\documentclass[fleqn,usenatbib]{mnras}
\usepackage{newtxtext,newtxmath}
\usepackage[T1]{fontenc}

\DeclareRobustCommand{\VAN}[3]{#2}
\let\VANthebibliography\thebibliography
\def\thebibliography{\DeclareRobustCommand{\VAN}[3]{##3}\VANthebibliography}

\usepackage{graphicx}	
\usepackage{amsmath}	
\usepackage[dvipsnames,svgnames]{xcolor}

\usepackage{color}

\title[BH Dynamics and Mergers]{Dynamical Friction Modeling of Massive Black Holes in Cosmological Simulations and Effects on Merger Rate Predictions}

\author[N.Chen et al.]{
Nianyi Chen,$^{1}$\thanks{E-mail: nianyic@andrew.cmu.edu}
Yueying Ni,$^{1}$
Michael Tremmel,$^{2}$
Tiziana Di Matteo,$^{1,3,4}$ Simeon Bird, $^5$
Colin DeGraf,$^{1}$
\newauthor
 Yu Feng $^6$
\\
$^{1}$McWilliams Center for Cosmology, Department of Physics, Carnegie Mellon University, Pittsburgh, PA 15213, USA\\
$^{2}$Astronomy Department, Yale University, P.O. Box 208120, New Haven, CT 06520, USA\\
$^3$NSF AI Planning Institute for Physics of the Future, 
Carnegie   Mellon  University, Pittsburgh, PA 15213, USA \\
$^{4}$ OzGrav-Melbourne, Australian Research Council Centre of Excellence for Gravitational Wave Discovery\\
$^{5}$ Department of Physics and Astronomy, University of California Riverside, Riverside, CA 90217, USA\\
$^{6}$Berkeley Center for Cosmological Physics and Department of Physics, University of California, Berkeley, CA 94720, USA
}

\date{Accepted XXX. Received YYY; in original form ZZZ}

\pubyear{2021}

\begin{document}
\label{firstpage}
\pagerange{\pageref{firstpage}--\pageref{lastpage}}
\maketitle

\begin{abstract}
In this work we establish and test methods for implementing dynamical friction for massive black hole pairs that form in large volume cosmological hydrodynamical simulations which include galaxy formation and black hole growth. We verify our models and parameters both for individual black hole dynamics and for the black hole population in cosmological volumes. Using our model of dynamical friction (DF) from collisionless particles, black holes can effectively sink close to the galaxy center, provided that the black hole's dynamical mass is at least twice that of the lowest mass resolution particles in the simulation. Gas drag also plays a role in assisting the black holes' orbital decay, but it is typically less effective than that from collisionless particles, especially after the first billion years of the black hole's evolution. DF from gas becomes less than $1\%$ of DF from collisionless particles for BH masses $> 10^{7}$ M$_{\odot}$.
Using our best DF model, we calculate the merger rate down to $z=1.1$ using an $L_{\rm box}=35$ Mpc$/h$ simulation box.  We predict $\sim 2$ mergers per year for $z>1.1$ peaking at $z\sim 2$.
These merger rates are within the range obtained in previous work using similar-resolution hydro-dynamical simulations. We show that the rate is enhanced by factor of $\sim 2$ when DF is taken into account in the simulations compared to the no-DF run. This is due to $>40\%$ more black holes reaching the center of their host halo when DF is added.

\end{abstract}



\section{Introduction}
\label{sec:introduction}
Super Massive Black Holes (SMBHs) are known to exist at the center of the majority of massive galaxies \citep[e.g.][]{Soltan1982,Kormendy1995,Magorrian1998,Kormendy2013}. 
As these galaxies merge \citep[e.g.][]{Lacey1993,Lotz2011,Rodriguez-Gomez2015}, the SMBHs that they host also go through mergers, resulting in the mass growth of the SMBH population \citep[e.g.][]{Begelman1980}.
SMBH mergers following their host galaxy mergers become an increasingly important aspect of SMBH growth for more massive black holes (BHs) in dense environments \citep[e.g.][]{Kulier2015}. 
As a by-product of BH mergers, gravitational waves are emitted, and their detection opens up a new channel for probing the formation and evolution of early BHs in the universe \citep[e.g.][]{Sesana2007a,Barausse2012}.

The gravitational wave detection by LIGO \citep[][]{LIGO2016PhRvL.116f1102A} proves the experimental feasibility of using gravitational waves for studying BH binaries. 
While LIGO cannot detect gravitational waves from binaries more massive than $\sim 100 M_\odot$ \citep[][]{Mangiagli2019}, long-baseline experiments are being planned for detections of more massive BH binaries. 
Specifically, the upcoming Laser Interferometer Space Antenna (LISA) \citep{LISA2017arXiv170200786A} mission will be sensitive to low-frequency ($10^{-4}-10^{-1}$Hz) gravitational waves from the coalescence of massive black holes (MBHs) with masses $10^4-10^7 M_\odot$ up to $z\sim 20$. 
 At even lower
frequencies Pulsar Timing Arrays (PTAs) are already collecting data and the
Square Kilometer Array (SKA) in the next decade will be a major leap forward in
sensitivity. PTA observations are likely to identify a number of
continuous-wave sources representing the early inspiral phase of MBHBs.
PTAs experiments \citep[e.g.][]{Jenet2004,Jenet2005} may also detect the inspiral of tight MBH binaries with mass $>10^8 M_\odot$. 
While massive BH binaries are the primary sources for PTAs and LISA, these
two experiments probe different stages of massive BH evolution. PTAs are most sensitive to the early inspiral
(orbital periods of years or longer) of nearby ($z <1$) (massive) sources  \citep{Mingarelli2017}. 
In contrast, LISA is sensitive to the inspiral, merger, and ringdown of
MBHBs  at a wide range of redshifts \citep{Amaro-Seoane2012}. 
The two populations of SMBHBs probed by PTAs and LISA are linked
via the growth and evolution of SMBH across cosmic time. 

LISA will provide a unique way of probing the high-redshift universe and understanding the early formation of the SMBHs, especially when combined with the soon-to-come observations of the electromagnetic (EM) counterparts \citep[][]{Natarajan2017,DeGraf2020}.
For instance, they will potentially allow us to distinguish between different BH seeding mechanisms at high-redshift \citep[][]{Ricarte2018}, to obtain information on the dynamical evolution of massive black holes \citep[][]{Bonetti2019}, and to gain information about the gas properties within the accretion disc \citep[][]{Derdzinski2019}. 

To properly analyze the upcoming results from the gravitational wave as well as the EM observations, we need to gain a thorough understanding of the physics of these MBH mergers with theoretical tools and be able to make statistical predictions on the binary population. In particular, it is important that the BH dynamics is modeled accurately, so that we can minimize the degeneracy with other physical properties of the merger, and gain accurate information about when and where BH coalescence is expected.

Hydrodynamical cosmological simulations provide a natural ground for studying the evolution and mergers of MBHs. In particular, large-volume cosmological simulations \citep[e.g.][]{Hirschmann2014,Vogelsberger2014,Schaye2015,Feng2016,Volonteri2016,Pillepich2018,Dave2019} have the statistical power to make merger rate predictions for the upcoming observations. 

In order to accurately predict when black hole mergers occur in these simulations, one must account for the long journey of the central black holes after the merger of their host galaxies: during galaxy mergers, the central SMBHs are usually separated by as much as a few tens of kpc. These SMBHs then gradually lose their orbital energy and sink to the center of the new galaxy due to the dynamical friction exerted by the gas, stars and dark matter around them \citep[e.g.][]{Chandrasekhar1943,Ostriker1999}. When their separation reaches the sub-parsec scale, they form a binary and other energy-loss channels begin to dominate, such as scattering with stars \citep[e.g.][]{Quinlan1996,Sesana2007b,Vasiliev2015}, gas drag from the circumbinary disk \citep[e.g.][]{Haiman2009}, or three-body scattering with a third black hole \citep[e.g.][]{Bonetti2018}.

However, due to limited mass and spatial resolution, large-scale cosmological simulations cannot feasibly include detailed treatment of the black hole binary dynamics. Without any additional correction to the BH dynamics, the smoothed-away small-scale gravity prevents effective orbital decay of the black hole after the orbit approaches the gravitational softening length. Once the binary reaches the innermost region of the remnant galaxy, the gravitational potential (close to the resolution limit) can be noisy. Such a noisy potential can scatter the black hole around within the host galaxy, or in some cases even kick the BH to the outskirts of the galaxy if the black hole mass is small. To avoid unexpected scattering of the BHs around the center of the galaxy, large-volume cosmological simulations usually resort to pinning the black holes at the halo minimum potential (a.k.a. repositioning). This repositioning algorithm has the undesirable effect of making the black holes merge rather efficiently
once they reach the center of the galaxy. Post-processing techniques have been used \citep[e.g.][]{Salcido2016,Kelley2017,Katz2020,Volonteri2020} to account for the additional dynamical friction effects on scales close to the gravitational smoothing scales of the black holes. This allows for an approximate estimation of the expected delay in the BH mergers. The post-processing calculations are mostly based on idealized analytical models, and therefore do not account for the variety of individual black hole environments.

Due to the increased merger efficiency induced by BH repositioning and the limits of post-processing in dynamical friction  calculations, emerging works have been adding sub-grid modeling of dynamical friction  self-consistently in cosmological simulations and removing the artificial repositioning approximation.
\cite{Chapon2013,Dubois2014} are the first large simulations to include the dynamical friction from gas, while \cite{Hirschmann2014} and \cite{Tremmel2017} account for dynamical friction from collisionless particles, and both have shown success in stabilizing the black holes at the halo centers. The dynamical friction modeling and its effect on the BH merger time scale have been well-tested in \cite{Tremmel2015} and \cite{Pfister2019} in the context of their relatively high-resolution simulations in a controlled single-halo environment, but they have also pointed out the failure of their model when the dark matter particle mass exceeds the black hole mass, and so their models might not be directly applicable to lower-resolution cosmological simulations. In the context of low-resolution cosmological simulations, the dynamical friction modeling is less well-tested, and its effects on the BH evolution and merger rate are not fully explored.

In this work, we carefully develop and test the sub-grid modeling of dynamical friction from both gas and collisionless particles in the context of cosmological simulations with resolution similar to the aforementioned large-volume, low-resolution hydrodynamical simulations (i.e. with a spatial resolution of $\sim$1kpc and mass resolution of $M_{\rm DM}\sim 10^7M_\odot$). We evaluate the models both by looking at individual black hole dynamics, growth and mergers, and by statistically comparing the behavior of different models in terms of the mass growth and merger statistics. In particular, we focus on how various models affect the BH merger rate in the cosmological simulations, which is essential for making merger rate predictions for the LISA mission.
    
This paper proceeds as follows: in Section \ref{sec:simulations} we describe the numerical code and the gaussian-constrained technique we use to study large SMBHs within a small volume. In Section \ref{sec:bh_model}, we talk about the different dynamical models for black hole mergers that we study and test in this work. Section \ref{sec:case} is dedicated to investigating the effect of the different models on the evolution of individual black holes, while Section \ref{sec:stats} studies the differences statistically. Finally, in Section \ref{sec:L35}, we show merger rate predictions with a model chosen based on the results of the previous sections, and compare with previous simulations at similar resolution.

\section{The Simulations}
\label{sec:simulations}

\subsection{The Numerical Code}
\label{subsec:code}

We use the massively parallel cosmological smoothed particle hydrodynamic (SPH) simulation software, MP-Gadget \citep{Feng2016}, to run all the simulations in this paper. 
The hydrodynamics solver of MP-Gadget adopts the new pressure-entropy formulation of SPH \citep{Hopkins2013}.
We apply a variety of sub-grid models to model the galaxy and black hole formation and associated feedback processes already validated against a number of observables \citep[e.g.][]{Feng2016,Wilkins2017,Waters2016,DiMatteo2017,Tenneti2018,Huang2018,Ni2018,Bhowmick2018,Marshall2020,Marshall2021}. Here we review briefly the main aspects of these.
In the simulations, gas is allowed to cool through radiative processes~\citep{Katz}, including metal cooling. For metal cooling, we follow the method in \cite{Vogelsberger2014}, and scale a solar metallicity template according to the metallicity of gas particles.
Our star formation (SF) is based on a multi-phase SF model ~\citep{SH03} with modifications following~\cite{Vogelsberger2013}.
We model the formation of molecular hydrogen and its effects on SF at low metallicity according to the prescription of \cite{Krumholtz}. 
We self-consistently estimate the fraction of molecular hydrogen gas from the baryon column density, which in turn couples the density gradient to the SF rate.
We include Type II supernova wind feedback ~\citep[the model used in BlueTides][]{Feng2016,Okamoto2010} in our simulations, assuming that the wind speed is proportional to the local one dimensional dark matter velocity dispersion.

BHs are seeded with an initial seed mass of $M_{\mathrm {seed}} = 5 \times 10^5 M_{\odot}/h$ in halos with mass more than $10^{10} M_{\odot}/h$ if the halo does not already contain a BH. We model BH growth and AGN feedback in the same way as in the \textit{MassiveBlack} $I \& II$ simulations, using the BH sub-grid model developed in \cite{SDH2005,DSH2005} with modifications consistent with BlueTides. 
The gas accretion rate onto the BHs is given by Bondi accretion rate,
\begin{equation}
\label{equation:Bondi}
    \dot{M}_B = \alpha \frac{4 \pi G^2 M_{\rm BH}^2 \rho}{(c^2_s+v_{\rm rel}^2)^{3/2}},
\end{equation}
where $c_s$ and $\rho$ are the local sound speed and density of the cold gas, $v_{\rm rel}$ is the relative velocity of the BH to the nearby gas, and $\alpha=100$ is a numerical correction factor introduced by \citep{Springel2005b}. This can also be eliminated (without affecting the values of the accretion rate significantly) in favor of a more detailed modeling of the contributions in the cold and hot phase accretion \citep{Pelupessy2006}.

We allow for super-Eddington accretion in the simulation \citep[e.g.][]{Volonteri2005,Volonteri2015}, but limit the accretion rate to 2 times the Eddington accretion rate:
\begin{equation}
\label{equation:Meddington}
    \dot{M}_{\rm Edd} = \frac{4 \pi G M_{\rm BH} m_p}{\eta \sigma_{T} c},
\end{equation}
where $m_p$ is the proton mass, $\sigma_T$ the Thompson cross section, c is the speed of light, and $\eta=0.1$ is the radiative efficiency of the accretion flow onto the BH.
Therefore, the BH accretion rate is determined by:
\begin{equation}
    \dot{M}_{\rm BH} = {\rm Min} (\dot{M}_{B}, 2\dot{M}_{\rm Edd}).
\end{equation}

The SMBH is assumed to radiate with a bolometric luminosity $L_{\rm Bol}$ proportional to the accretion rate $\dot{M}_{\rm BH}$:
\begin{equation}
    L_{\rm Bol} = \eta \dot{M}_{\rm BH} c^2
\end{equation}
with $\eta = 0.1$ being the mass-to-light conversion efficiency in an accretion disk according to \cite{Shakura1973}.
5\% of the radiated energy is thermally coupled to the surrounding gas that resides within twice the radius of the SPH smoothing kernel of the BH particle. This scale is typically about $\sim 1-3\%$ of the virial radius of the halo.

The cosmological parameters used are from the nine-year Wilkinson Microwave Anisotropy Probe (WMAP) \citep{Hinshaw2013} ($\Omega_0=0.2814$, $\Omega_\Lambda=0.7186$, $\Omega_b=0.0464$, $\sigma_8=0.82$, $h=0.697$, $n_s=0.971$).
For our fiducial resolution simulations, the mass resolution is $M_{\rm DM} = 1.2 \times 10^7 M_\odot/h$ and $M_{\rm gas} = 2.4 \times 10^6 M_\odot/h$ in the initial conditions.
The mass of a star particle is $M_{*} = 1/4 M_{\rm gas} = 6 \times 10^5 M_\odot/h$. The gravitational softening length is $\epsilon_g = 1.5$ ckpc/$h$ in the fiducial resolution for both DM and gas particles. The detailed simulation and model parameters are listed in Tables \ref{tab:cons} and \ref{tab:norm}. 
\subsection{Gaussian Constrained Realization}
\label{subsec:CR}

MBHs at high redshift typically reside in rare density peaks, which are absent in the small uniform box ($\sim 10$ Mpc/$h$) simulations. 
In order to test the dynamics for more massive BHs (with $M_{\rm BH} > 10^8 M_{\odot}$) in our small volume simulation, we apply the Constrained Realization (CR) technique \footnote{\url{https://github.com/yueyingn/gaussianCR}} to impose a relatively high density peak in the initial condition (IC), with peak height $\nu = 4 \sigma_0$ on scale of $R_G = 1$ Mpc/$h$.  

The prescription for the CR technique was first introduced by \cite{Hoffman1991} as an optimal way to construct samples of constrained Gaussian random fields.
This formalism was further elaborated and extended by \cite{vandeWeygaert1996} as a more general type of convolution format constraints.
The CR technique imposes constraints on different characteristics of the linear density field. 
It can specify density peaks in the Gaussian random field with any desired height and shape, providing an efficient way to study rare massive objects with a relatively small box and thus lower computational costs \citep[e.g.][]{Ni2020}.
In this study, we specify a $4 \sigma_0$ density peak in the IC of our $10$ Mpc/$h$ box, boosting the early formation of halos and BHs to study the dynamics of massive BHs. Before applying the peak height constraint, the highest density peak has $\nu = 2.4 \sigma_0$ and the largest BH has mass $<6\times 10^7M_\odot$ at $z=3$ in our fiducial model (\texttt{DF\_4DM\_G} in Table \ref{tab:cons}). After applying the $4 \sigma_0$ constraint, the largest BH has mass $3\times 10^8 M_\odot$ at $z=3$ in the same box.

\begin{table*}
\caption{Constrained Simulations}
\label{tab:cons}
\begin{tabular}{lccccccc}
\hline
Name & Lbox & ${\rm N}_{\rm part}$ & ${\rm M}_{\rm DM}$ & ${\rm M}_{\rm Dyn,seed}$ & $\epsilon_{\rm g}$ & BH Dynamics & Merging Criterion \\
& [$h^{-1}$Mpc] & & [$h^{-1} {\rm M}_\odot$] & [${\rm M}_{\rm DM}$] & [$h^{-1}{\rm kpc}$] & &  \\
\hline
NoDF\_4DM & 10 & $176^3$ & $1.2\times 10^7$ & 4 & 1.5 & gravity & distance\\
NoDF\_4DM\_G & 10 & $176^3$ & $1.2\times 10^7$ & 4 & 1.5 & gravity & distance \& grav.bound\\
DF\_4DM & 10 & $176^3$ & $1.2\times 10^7$ & 4 & 1.5 & gravity+DF & distance\\
Drag\_4DM\_G & 10 & $176^3$ & $1.2\times 10^7$ & 4 & 1.5 & gravity+Drag & distance \& grav.bound\\
DF+Drag\_4DM\_G & 10 & $176^3$ & $1.2\times 10^7$ & 4 & 1.5 & gravity+DF+Drag & distance \& grav.bound\\
DF\_4DM\_G & 10 & $176^3$ & $1.2\times 10^7$ & 4 & 1.5 & gravity+DF & distance \& grav.bound\\
DF\_2DM\_G & 10 & $176^3$ & $1.2\times 10^7$ & 2 & 1.5 & gravity+DF & distance \& grav.bound\\
DF\_1DM\_G & 10 & $176^3$ & $1.2\times 10^7$ & 1 & 1.5 & gravity+DF & distance \& grav.bound\\
DF(T15)\_4DM\_G & 10 & $176^3$ & $1.2\times 10^7$ & 4 & 1.5 & gravity+DF(T15) & distance \& grav.bound\\
DF\_HR\_4DM\_G & 10 & $256^3$ & $4\times 10^6$ & 4 & 1.0 & gravity+DF & distance \& grav.bound\\
DF\_HR\_12DM\_G & 10 & $256^3$ & $4\times 10^6$ & 12 & 1.0 & gravity+DF & distance \& grav.bound\\
\hline
\end{tabular}
\end{table*}

\begin{table*}
\caption{Unconstrained Simulations}
\label{tab:norm}
\begin{tabular}{lccccccc}
\hline
Name & Lbox & ${\rm N}_{\rm part}$ & ${\rm M}_{\rm DM}$ & ${\rm M}_{\rm Dyn,seed}$ & $\epsilon_{\rm g}$ & BH Dynamics & Merging Criterion \\
& [$h^{-1}$Mpc] & & [$h^{-1} {\rm M}_\odot$] & [${\rm M}_{\rm DM}$] & [$h^{-1}{\rm kpc}$] & &  \\
\hline
L15\_Repos\_4DM & 15 & $256^3$ & $1.2\times 10^7$ & 4 & 1.5 & reposition & distance\\
L15\_NoDF\_4DM & 15 & $256^3$ & $1.2\times 10^7$ & 4 & 1.5 & gravity & distance\\
L15\_NoDF\_4DM\_G & 15 & $256^3$ & $1.2\times 10^7$ & 4 & 1.5 & gravity & distance \& grav.bound\\
L15\_DF\_4DM & 15 & $256^3$ & $1.2\times 10^7$ & 4 & 1.5 & gravity+DF & distance\\
L15\_DF\_4DM\_G & 15 & $256^3$ & $1.2\times 10^7$ & 4 & 1.5 & gravity+DF & distance \& grav.bound\\
L15\_DF(T15)\_4DM\_G & 15 & $256^3$ & $1.2\times 10^7$ & 4 & 1.5 & gravity+DF(T15) & distance \& grav.bound\\
L15\_DF+drag\_4DM\_G & 15 & $256^3$ & $1.2\times 10^7$ & 4 & 1.5 & gravity+DF+Drag & distance \& grav.bound\\
L35\_NoDF\_4DM\_G & 35 & $600^3$ & $1.2\times 10^7$ & 4 & 1.5 & gravity+DF & distance \& grav.bound\\
L35\_DF+drag\_4DM\_G & 35 & $600^3$ & $1.2\times 10^7$ & 4 & 1.5 & gravity+DF+drag & distance \& grav.bound\\

\hline
\end{tabular}
\end{table*}
\section{BH Dynamics}
\label{sec:bh_model}
\subsection{BH Dynamical Mass}
\label{subsec:mdyn}
In our simulations, the seed mass of the black holes is $5\times 10^5 M_\odot/h$, which is 20 times smaller than the fiducial dark matter particle mass at $1.2\times 10^7 M_\odot/h$. Such a small mass of the BH relative to the dark matter particles will result in very noisy gravitational acceleration on the black holes, and causes instability in the black hole's motion as well as drift from the halo center. Moreover, as shown in previous works \citep[e.g.][]{Tremmel2015,Pfister2019}, under the low $M_{\rm BH}/M_{\rm DM}$ regime, it is challenging to effectively model dynamical friction in a sub-grid fashion.

To alleviate dynamical heating by the noisy potential due to the low $M_{\rm BH}/M_{\rm DM}$ ratio, we introduce a second mass tracer, the dynamical mass $M_{\rm dyn}$, which is set to be comparable to $M_{\rm DM}$ when the black hole is seeded. This mass is used in force calculation for the black holes, including the gravitational force and dynamical friction, while the intrinsic black hole mass $M_{\rm BH}$ is used in the accretion and feedback process. $M_{\rm dyn}$ is kept at its seeding value $M_{\rm dyn,seed}$ until $M_{\rm BH}>M_{\rm dyn,seed}$. After that $M_{\rm dyn}$ grows following the black hole's mass accretion. With the boost in the seed dynamical mass, the sinking time scale will be shortened by a factor of $\sim M_{\rm BH}/M_{\rm dyn}$ compared to the no-boost case. Note that the bare black hole sinking time scale estimated in the no-boost case could over-estimate the true sinking time, as the high-density stellar bulges sinking together with the black hole are not fully resolved \citep[e.g.][]{Antonini2012,Dosopoulou2017,Biernacki2017}.

 The boost we need to prevent dynamical heating depends on the dark matter particle mass $M_{\rm DM}$ (if we have high enough resolution the boost is no longer necessary), so we parametrize the dynamical mass in terms of the dark matter particle mass, $M_{\rm dyn,seed} = k_{\rm dyn} M_{\rm DM}$, instead of setting an absolute seeding dynamical mass for all simulations. We expect that as we go to higher resolutions where $M_{\rm DM}$ is comparable to $M_{\rm BH,seed}$, the dynamical seed mass should converge to the black hole seed mass, if we keep $k_{\rm dyn}$ constant. We study the effect of setting different $k_{\rm dyn}$ by running three simulations with the same resolution and dynamical friction models, but various $k_{\rm dyn}$ ratios. They are listed in Table \ref{tab:cons} as \texttt{DF\_4DM\_G}, \texttt{DF\_2DM\_G}, and \texttt{DF\_1DM\_G}, with $k_{\rm dyn}=4,2,1$, respectively.

To explore the effects of the BH seed dynamical mass on the motion and mergers of the black hole, we test a variety of $M_{\rm dyn,seed}$ values in our simulations. The comparison between different $M_{\rm dyn,seed}$ can be found in Appendix \ref{app:res}. 
\subsection{Modeling of Black Hole Dynamics}
\label{subsec:df}
\subsubsection{Reposition of the Black Hole}

Before introducing our dynamical friction implementations, we first describe a baseline model utilized by many large-volume cosmological simulations: the reposition model. As the name suggests, the reposition model of black hole dynamics places the black hole at the location of a local gas particle with minimum gravitational potential at each time step, in order to avoid the unrealistic motion of the black holes due to limited mass and force resolution. This is particularly preferred for large-volume, low-resolution cosmological simulations \citep[e.g.][]{Springel2005b, Sijacki2007, Booth2009,Schaye2015,Pillepich2018}, where the black hole mass is smaller than a star or gas particle mass and the BH can be inappropriately scattered around by two-body forces as well as the noisy local potential.

This simple fix of repositioning, however, comes with many disadvantages. For example, it may lead to higher accretion and feedback of the black holes, as they sink to the high-density regions too quickly. As was shown in \cite{Wurster2013} and \cite{Tremmel2017}, repositioning also leads to burstier feedback of the BHs, which is more likely to quench star-formation in the host galaxies. Moreover, repositioning leads to ill-defined velocity and non-smooth trajectories of the black hole particles. Because of the ill-defined velocity and extremely short orbital decay time, such methods cannot be reliably used for merger rate predictions without careful post-processing calculations to account for the orbital decays.

In our study, we use the reposition model as a reference for the black hole statistics, as it is still widely adopted in many existing simulations. We want to compare the dynamical friction models with the reposition model and quantify the effect of repositioning on BH mass growth and merger rate compared with the dynamical friction models.

\subsubsection{Dynamical Friction from Collisionless Particles}

When the black hole travels through a continuous medium or a medium consisting of particles with smaller masses than the black hole, it attracts the surrounding mass towards itself, leaving a tail of overdensity behind.  Dynamical friction is the resulting gravitational force exerted onto the black hole by this tail of overdensity \citep[e.g.][]{Chandrasekhar1943,Binney2008}. Dynamical friction causes the orbits of SMBHs to decay towards the center of massive galaxies \citep[e.g.][]{Governato1994,Kazantzidis2005}, and enables the black holes to stay at the high-density regions where they could go through efficient accretion and mergers.

We follow Equation (8.3) in \cite{Binney2008} for the acceleration of the black hole due to dynamical friction:

\begin{equation}
\label{eq:df_full}
    \mathbf{F}_{\rm DF} = -16\pi^2 G^2 M_{\rm BH}^2 m_{a} \;\text{log}(\Lambda) \frac{\mathbf{v}_{\rm BH}}{v_{\rm BH}^3} \int_0^{v_{\rm BH}} dv_a v_a^2 f(v_a),
\end{equation}
where $M_{\rm BH}$ is the black hole mass, $\textbf{v}_{\rm BH}$ is the velocity of the black hole relative to its surrounding medium, $m_a$ and $v_a$ are the masses and velocities of the particles surrounding the black hole, and $\text{log}(\Lambda)=\text{log}(b_{\rm max}/b_{\rm min})$ is the Coulomb logarithm that accounts for the effective range of the friction between $b_{\rm min}$ and $b_{\rm max}$(we will specify how we set these parameters later). $f(v_a)$ is the velocity distribution of the surrounding particles (unless we explicitly state otherwise, all variables involving the black hole's surrounding particles are calculated using stars and dark matter particles). Here we have assumed an isotropic velocity distribution of the particles surrounding the black hole, so that we are left with an 1D integration. 

We test two different numerical implementations of the dynamical friction (DF) in our simulations: one with a more aggressive approach which likely overestimates the effective range of DF, but could be more suitable for large-volume simulations (we refer to it as DF(fid) in places where we carry out explicit comparisons between the two DF models, and drop the 'fid' in all other places); the other with a more conservative method which aims to only account for the DF below the gravitational softening length, and is well-tested for smaller volume, high-resolution simulations \citep{Tremmel2015} (we refer to it as DF(T15)).

We begin by introducing the DF(fid) model. In this model, we further follow the derivation in \cite{Binney2008}, and approximate $f(v_a)$ by the Maxwellian distribution, so that Equation \ref{eq:df_full} reduces to:
\begin{equation}
    \label{eq:H14}
    \mathbf{F}_{\rm DF,fid} = -4\pi \rho_{\rm sph} \left(\frac{GM_{\rm dyn}}{v_{\rm BH}}\right)^2  \;\text{log}(\Lambda_{\rm fid}) \mathcal{F}\left(\frac{v_{\rm BH}}{\sigma_v}\right) \frac{\bf{v}_{\rm BH}}{v_{\rm BH}}.
\end{equation}
Here $\rho_{\rm sph}$ is the density of dark matter and star particles within the SPH kernel (we will sometimes refer to these particles as "surrounding particles") of the black hole. All other definitions follow those of Equation \ref{eq:df_full}, except that we have substituted $M_{\rm BH}$ with $M_{\rm dyn}$ following the discussion in \ref{subsec:mdyn}.
The function $\mathcal{F}$ defined as:
\begin{equation}
    \label{eq:fx}
    \mathcal{F}(x) =  \text{erf}(x)-\frac{2x}{\sqrt{\pi}} e^{-x^2}, \;
    x=\frac{v_{\rm BH}}{\sigma_v}
\end{equation}
is the result of analytically integrating the Maxwellian distribution, where $\sigma_v$ is the velocity dispersion of the surrounding particles.

The subscript "fid" in $\text{log}(\Lambda)$ means that this definition of $\Lambda$ is specific to the DF(fid) model, with
\begin{equation}
    \Lambda_{\rm fid} = \frac{b_{\rm max,fid}}{(GM_{\rm dyn})/v_{\rm BH}^2}, \; b_{\rm max,fid} = 10\text{ ckpc}/h.
\end{equation}
Note that here we have defined $b_{\rm max}$ as a constant roughly equal to 6 times the gravitational softening. As there is no general agreement on the distance above which dynamical friction is fully resolved, we tested several values ranging from $\epsilon_g$ to $20\epsilon_g$. We found that values above $2\epsilon_g$ are effective in sinking the black hole, although a smaller $b_{\rm max}$ tends to result in more drifting black holes at higher redshift. By using this definition, we are likely overestimating the effective range of dynamical friction. However, we find this over-estimation necessary in the early stage of black hole growth to stabilize the black hole motion.

We also implement a more localized version of dynamical friction following  \cite{Tremmel2015} which we call DF(T15). Under the DF(T15) model, the dynamical friction is expressed as:

\begin{equation}
    \label{eq:T15}
    \mathbf{F}_{\rm DF,T15} = -4\pi \rho (v<v_{\rm BH}) \left(\frac{GM_{\rm dyn}}{v_{\rm BH}}\right)^2  \text{log}(\Lambda_{\rm T15}) \frac{\bf{v}_{\rm BH}}{v_{\rm BH}}.
\end{equation}
Here the surrounding density only accounts for the particles moving slower than the BH with respect to the environment. More formally,
\begin{equation}
\label{eq:rho}
    \rho (v<v_{\rm BH}) = \frac{M(<v_{\rm BH})}{M_{\rm total}} \rho_{\rm T15},
\end{equation}
where $M_{\rm total}$ is the total mass of the nearest 100 DM and stars, $M(<v_{\rm BH})$ is the fractional mass counting only DM and star particles with velocities smaller than the BH, and $\rho_{T15}$ is the density calculated from the nearest 100 DM/Star particles (note that in comparison, the SPH kernel contains 113 gas particles but far more collisionless particles (see Figure \ref{fig:k100_case1})). By using $\rho (v<v_{\rm BH})$ in place of $\rho_{\rm sph} \mathcal{F}$, we are approximating the velocity distribution of surrounding particles by the distribution of the nearest 100 collisionless particles. Another major difference from the DFsph model is the Coulomb logarithm, where in this model we define:
\begin{equation}
    \Lambda_{\rm T15} = \frac{b_{\rm max,T15}}{(GM_{\rm dyn})/v_{\rm BH}^2}, \; b_{\rm max,T15} = \epsilon_g.
\end{equation}
The choice of a lower $b_{\rm max}$ is consistent with the localized density and velocity calculations, and by doing so we have assumed that dynamical friction is fully resolved above the gravitational softening.

\subsubsection{Gas Drag}
\label{subsection:drag}
In addition to the dynamical friction from dark matter and stars, the black hole can also lose its orbital energy due to the dynamical friction from gas (to distinguish from dynamical friction from dark matter and stars, we will refer to the gas dynamical friction as "gas drag" hereafter). \cite{Ostriker1999} first came up with the analytical expression for the gas drag term from linear perturbation theory, and showed that in the transonic regime the gas drag can be more effective than the dynamical friction from collisionless particles. Although later studies show that \cite{Ostriker1999} likely overestimates the gas drag for gas with Mach numbers slightly above unity \citep[e.g.][]{Escala2004ApJ,Chapon2013}, simulations with gas drag implemented still demonstrate that this is an effective channel for black hole energy loss during orbital decays \citep[e.g.][]{Chapon2013,Dubois2013,Pfister2019}.

In order to investigate the relative effectiveness of DF and gas drag, we also include gas drag onto black holes in our simulations following the analytical approximation from \cite{Ostriker1999}:
\begin{equation}
\label{eq:drag}
    \mathbf{F}_{\rm drag} = -4 \pi\rho \left( \frac{G M_{\rm dyn}}{c_s^2} \right)^2 \times \mathcal{I(M)}\frac{\bf{v}_{\rm BH}}{v_{\rm BH}},
\end{equation}
where $c_s$ is the sound speed, $\mathcal{M} = \frac{| \mathbf{v}_{\rm BH} - \mathbf{v}_{\rm gas}|}{c_s}$ is the Mach number, and $\mathcal{I(M)}$ is given by:
\begin{align}
    \mathcal{I}_{\rm subsonic} &= \mathcal{M}^{-2} \left[ \frac{1}{2} \text{log}\left(\frac{1+\mathcal{M}}{1-\mathcal{M}}\right) -\mathcal{M}\right] \\
    \mathcal{I}_{\rm supersonic} &= \mathcal{M}^{-2} \left[ \frac{1}{2} \text{log}\left(\frac{\mathcal{M}+1}{\mathcal{M}-1}\right) -\text{log} \Lambda_{\rm fid} \right],
\end{align}
where $\text{log} \Lambda_{\rm fid}$ is the Coulomb logarithm defined similarly to the collisionless dynamical friction.

\begin{figure*}
\includegraphics[width=0.49\textwidth]{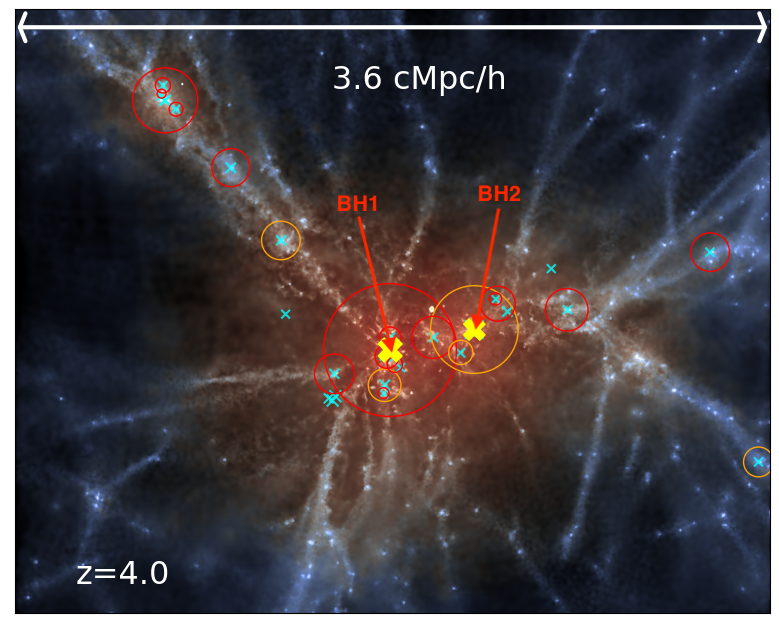}
\includegraphics[width=0.49\textwidth]{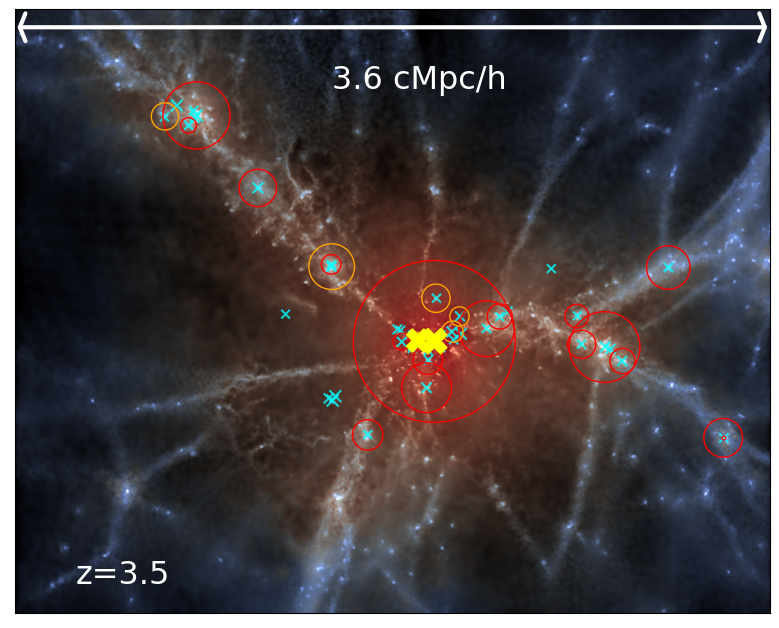}

\caption{Visualization of $4\sigma_0$ density peak of the \texttt{DF\_4\_DM\_G} simulation at $z=4.0$ and $z=3.5$. The brightness corresponds to the gas density, and the warmness of the tone indicates the mass-weighted temperature of the gas. We plot the black holes (\textbf{cross}) with mass $>10^6 M_\odot$, as well as the halos (subhalos) hosting them (\textbf{red circles} correspond to central halos, \textbf{orange circles} correspond to subhalos. The circle radius shows the virial radius of the halo; halos are identified by Amiga's Halo Finder(AHF)). This density peak hosts the two largest black holes in our simulations (\textbf{yellow cross}), and they are going through a merger along with the merger of their host halos between $z=4$ and $z=3$. For the black hole and merger case studies, we will use examples from the circled halos/black holes shown in this figure.} 
\label{fig:halos}
\end{figure*}

\subsection{Merging Criterion}
\label{subsec:merger}
In all of our simulations, we set the merging distance to be $2\epsilon_{\rm g}$, because the BH dynamics below this distance is not well-resolved due to our limited spatial resolution. We conserve the total momentum of the binary during the merger.

Under the baseline repositioning treatment of the BH dynamics, the velocity of the black hole is not a well-defined quantity. Therefore, in cosmological simulations with repositioning, the distance between the two black holes is often the only criterion imposed during the time of mergers (for example BlueTides \citep{Feng2016}, Illustris \citep{Vogelsberger2013} and IllustrisTNG \citep{Pillepich2018}). One problem with using only the distance as a merging criterion is that it can spuriously merge two passing-by black holes with high velocities, when in reality they are not gravitationally bound and should not merge just yet (or may never merge). Although some similar-resolution simulations such as EAGLE \citep{Crain2015,Schaye2015} also check whether two black hole particles are gravitationally bound, the black holes still do not have a well-defined orbit and sinking time due to the discrete positioning.

When we turn off the repositioning of the BHs to the nearby minimum potential, the BHs will have well-defined velocities at each time step (this is true whether or not we add the dynamical friction). This allows us to apply further merging criteria based on the velocities and accelerations of the black hole pair, and thus avoid earlier mergers of the gravitationally unbound pairs. Also, as the BH pairs now have well-defined orbits all the way down to the numerical merger time, we will be able to directly measure binary separation and eccentricity from the numerical merger, and use the measurements as the initial condition for post-processing methods without having to assume a constant initial value \citep[e.g.][]{Kelley2017}.

We follow \cite{Bellovary2011} and \cite{Tremmel2017}, and use the criterion
\begin{equation}
    \label{eq:merge_criterion}
    \frac{1}{2}|\bf{\Delta v}|^2 < \bf{\Delta a} \bf{\Delta r}
\end{equation}
 to check whether two black holes are gravitationally bound. Here $\bf{\Delta a}$,$\bf{\Delta v}$ and $\bf{\Delta r}$ denote the relative acceleration, velocity and position of the black hole pair, respectively. Note that this expression is not strictly the total energy of the black hole pair, but an approximation of the kinetic energy and the work needed to get the black holes to merge. Because in the simulations the black hole is constantly interacting with surrounding particles, on the right-hand side we use the overall gravitational acceleration instead of the acceleration purely from the two-body interaction.

\section{Case Studies of BH Models}
\label{sec:case}

\begin{figure*}
\includegraphics[width=0.91\textwidth]{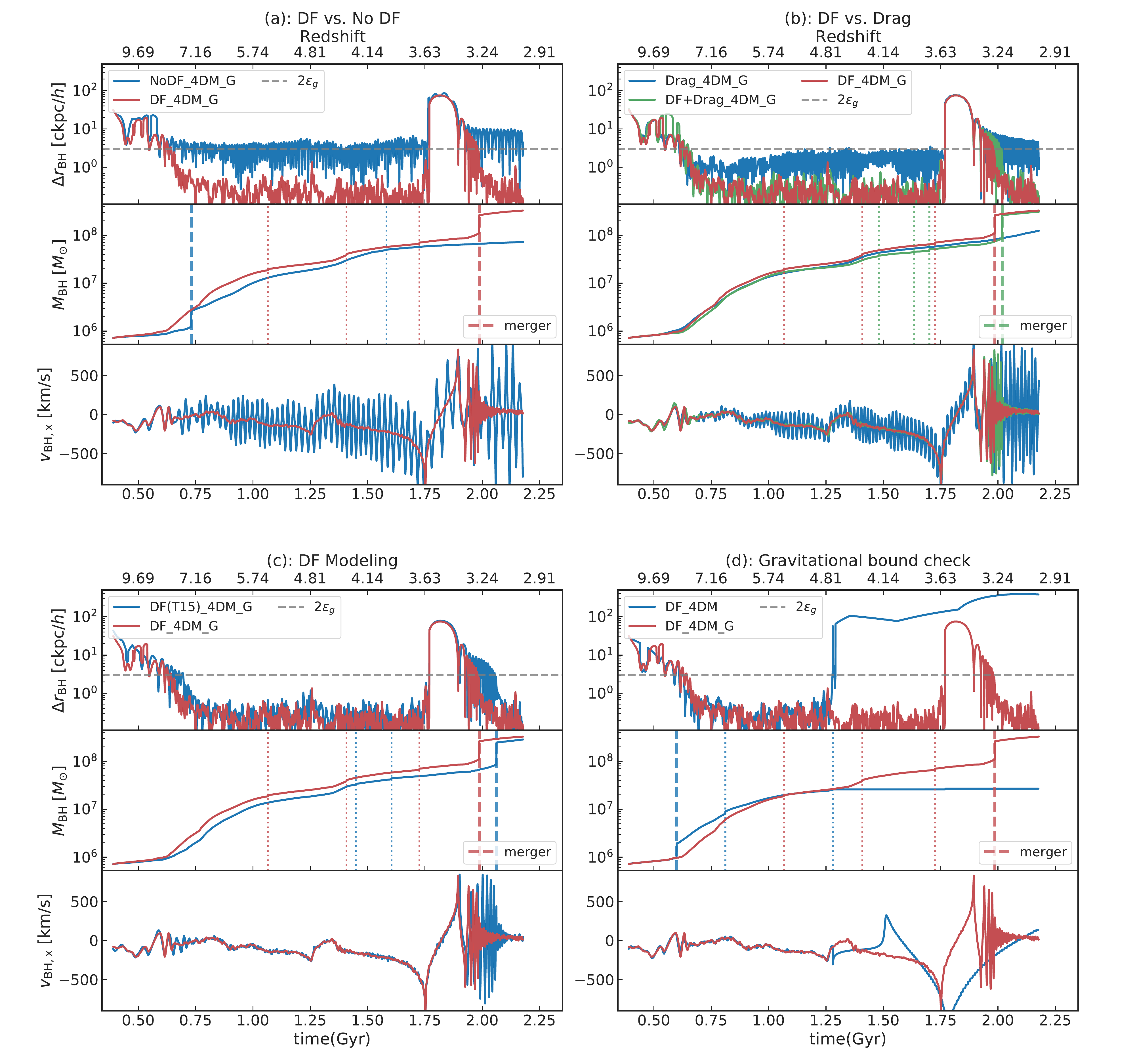}

\caption{ The evolution of BH2 in Figure \ref{fig:halos} under different BH dynamics prescriptions. We show the distance to halo center (\textbf{top}), black hole mass (\textbf{middel}) and the $x$-component of the black hole velocity (\textbf{bottom}). Mergers are shown in vertical lines (thick dashed lines are major mergers ($q>0.3$), and thin dotted lines are minor mergers) \textbf{(a):} comparison between no-DF and DF models. DF clearly helps the black hole sink to the halo center and stay there. \textbf{(b):} Effects of DF from stars and dark matter compared with gas drag. DF has a stronger effect throughout, except that in the very early stage the drag-only model is comparable to the DF-only model. \textbf{(c)}: Comparison between the DF(fid) and DF(T15) model. In general, the DF(fid) model results in a more stable black hole motion and faster sinking, but the difference is small. \textbf{(d)}: Black hole dynamics with and without the gravitational bound check during mergers. Without the gravitational bound check, the black holes can merge while still moving with large momenta, and thereby get kicked out of the halo by the injected momentum.}
\label{fig:big_plot}
\end{figure*}


Given the variety of models we have described so far, we first study the effect of different BH dynamics models by looking at the individual black hole evolution and black hole pairs using the constrained simulations. The details of these simulations and specific dynamical models are shown in Table \ref{tab:cons}. For all the constrained simulations, we use the same initial conditions, which enables us to do a case-by-case comparison between different BH dynamical models.

For the case studies, we choose to study the growth and merger histories of the two largest black holes and a few surrounding black holes within the density peak of our simulations. The halos and black holes at the $4\sigma_0$ density peak in \texttt{DF\_4DM\_G} are shown in Figure \ref{fig:halos}. The halos and subhalos shown in circles are identified with Amiga's Halo Finder \citep[AHF,][]{Knollmann2009}. The halos are centered at the minimum-potential gas particle within the halo, and the sizes of the circles correspond to the virial radius of the halo. Throughout the paper, we will always define the halo centers by the position of the minimum-potential gas particle, and we note that the offset between the minimum-potential gas and the halo center given by AHF (found via density peaks) is always less than 1.5 ckpc$/h$. The cyan crosses are black holes with mass larger than $10^6 M_\odot/h$, and the yellow crosses are the two largest black holes in the simulation. From the plot, we can see that in the \texttt{DF\_4DM\_G} simulation, most of the black holes already reside in the center of their hosting halos at $z=4$, although we also see some cases of wandering BHs outside of the halos.

\subsection{Black Hole Dynamics Modeling}
\label{subsec:models}

To compare different dynamical models, we look at the distance between the black hole and the halo center $\Delta r_{\rm BH}$ (we will sometimes refer to this distance as "drift" hereafter), the black hole mass, and the velocity along the $x$ direction through the entire history of BH2 from Figure \ref{fig:halos}. 
 
 We evaluate the black hole drift with two approaches: at each time-step, we find the minimum potential gas particle within 10 ckpc$/h$ of the black hole and calculate the distance between this gas particle and the black hole. This is a quick evaluation of the drift that allows us to trace the black hole motion at each time step, but it fails to account for orbits larger than 10 ckpc$/h$, and the minimum-potential gas particle may not reside in the same halo as the black hole. Therefore, for each snapshot we saved, we define the drift more carefully by running the halo finder and calculate the distance between the black hole and the center of its host halo. Whenever the black hole is further than 9 ckpc$/h$ from the minimum potential gas particle, we take the distance from the two nearest snapshots and linearly interpolate in time between them. Otherwise we use the 
 distance to the local minimum potential gas particle calculated at each time step.

\subsubsection{DF and No Correction}

Before calibrating our dynamical friction modeling, we first demonstrate the effectiveness of our fiducial DF model, \texttt{DF\_4DM\_G}, by comparing it with the no-DF run \texttt{NoDF\_4DM\_G} (note that throughout the paper, no-DF means no correction to the BH dynamics of any form besides the resolved gravity). We keep all parameters fixed except for the black hole dynamics modeling. The details of these simulations can be found in Table \ref{tab:cons}.

In Figure \ref{fig:big_plot}(a), we show the evolution of BH2 in Figure \ref{fig:halos} under the no-DF and the fiducial DF models. Without any correction to the black hole dynamics, even the largest black hole in the simulation does not exhibit efficient orbital decay throughout its evolution: the distance from the halo center is always fluctuating above $2\epsilon_g$. This is because the black hole does not experience enough gravity on scales below the softening length, and cannot lose its angular momentum efficiently. Now when we add the additional dynamical friction to compensate for the missing small-scale gravity, the black hole is able to sink to within 1 ckpc$/h$ of the halo centers in <200 Myr and remain there. 

The 90 ckpc$/h$ peak in the drift of the black hole marks the merger between BH1 and BH2 in Figure \ref{fig:halos}, when the host halo of BH2 merges into the host of BH1, and the halo center is redefined near the merger. After the halo merger, dynamical friction is able to sink the black hole to the new halo center and allows it to merge with the black hole in the other halo, whereas in the no-DF case we do not see the clear orbital decay of the black holes after the merger of their host halo until the end of the simulation.

Besides the drift, we also show the x-component of the black hole's velocity relative to its surrounding collisionless particles (lower panel). Here we show one component instead of the magnitude to better visualize the velocity oscillation. With dynamical friction turned on, the velocity of the black hole is more stable, as the black hole's orbit has already become small and is effectively moving together with the host halo. Without dynamical friction, the black hole tends to oscillate with large velocities around the halo center without losing its angular momentum.

The different dynamics of the black hole can also affect accretion due to differences in density and velocities, so we also look at the black holes' mass growth in the two scenarios (middle panel). The mass growths of the two black holes are similar under the two models, although when subjected to dynamical friction, the black holes have more and earlier mergers. Even though the black hole mass is less sensitive to the dynamics modeling, the merger rate predictions can be affected significantly as we will discuss later. 

Note that for our no-DF model, we have also boosted the dynamical mass to $4\times M_{\rm DM}$ at the early stage to prevent scattering by the dark matter and star particles. However, even after the boost, the black holes cannot lose enough angular momentum to be able to stay at the halo center. This means that even though dynamical heating is alleviated through the large dynamical mass, the sub-resolution gravity is still essential in sinking the black hole to the host halo center.


\subsubsection{Dynamical Friction and Gas Drag}
\label{subsec:drag}
\begin{figure*}
\includegraphics[width=0.49\textwidth]{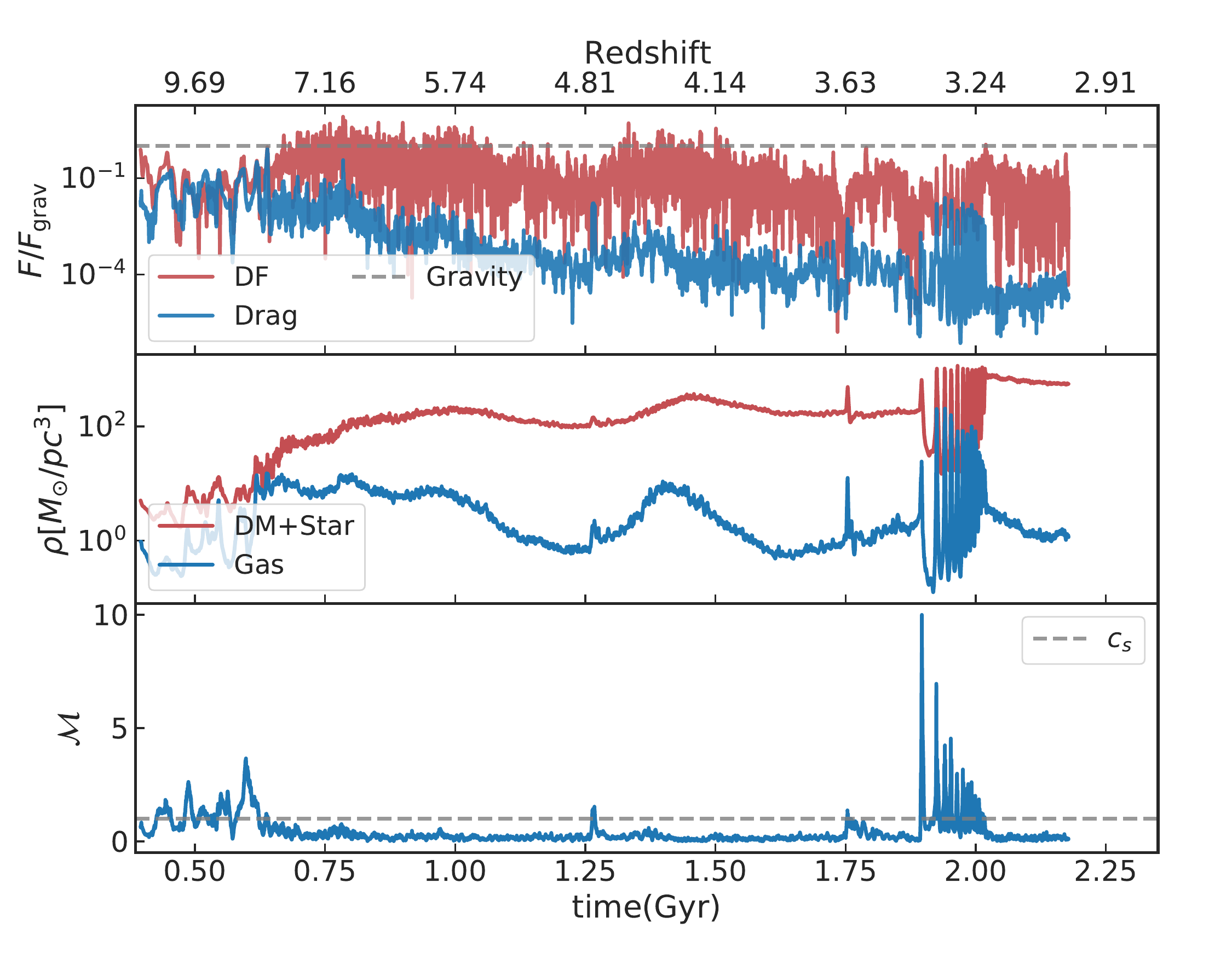}
\includegraphics[width=0.49\textwidth]{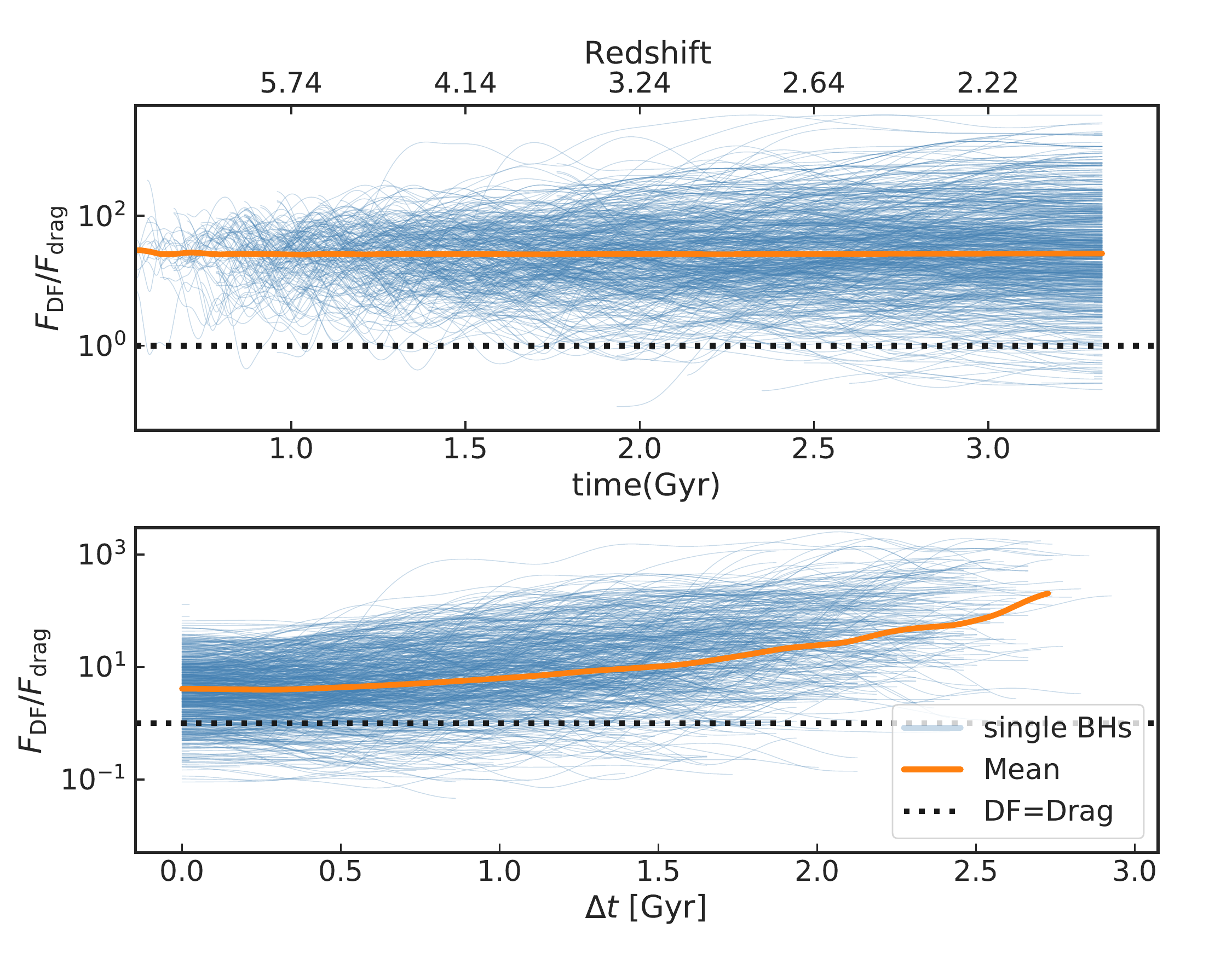}

\caption{Comparisons between DF and hydro drag. \textbf{Left:} comparison for a single black hole. In the top panel we show the magnitude of the DF (\textbf{red}) and gas drag (\textbf{blue}) relative to gravity for the same black hole, in the \texttt{DF+Drag\_4DM\_G} run. During the early stage of the black hole evolution, DF and gas drag have comparable effect, while after $z=7.5$ the gas drag becomes less and less important, as the gas density decreases relative to the stellar density (\textbf{middle}), and the black hole velocity goes into the subsonic regime (\textbf{lower}). \textbf{Right:} Ratio between DF and gas drag for all black holes. We plot the ratio both as a function of redshift (\textbf{top}) and as a function of time after a black hole is seeded (\textbf{bottom}). The orange lines represent the logarithmic mean of the scatter. The $F_{\rm DF}/F_{\rm drag}$ ratio depends strongly on the evolution time of the black hole: the longer the black hole evolves, the less important the drag force is. However, there is not a strong correlation between redshift and the $F_{\rm DF}/F_{\rm drag}$ ratio.}
\label{fig:drag}
\end{figure*}

\begin{figure*}
\includegraphics[width=0.33\textwidth]{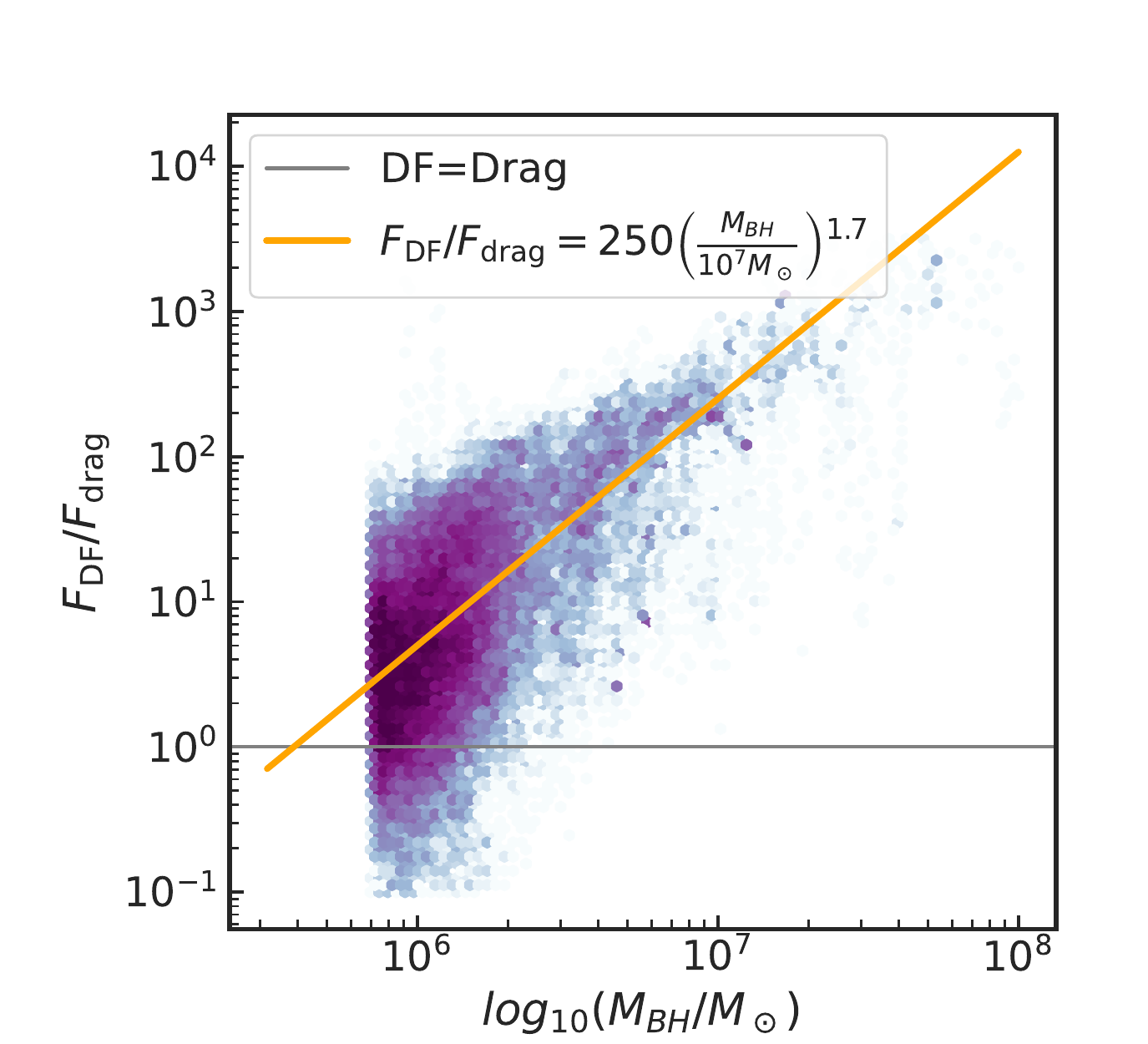}
\includegraphics[width=0.66\textwidth]{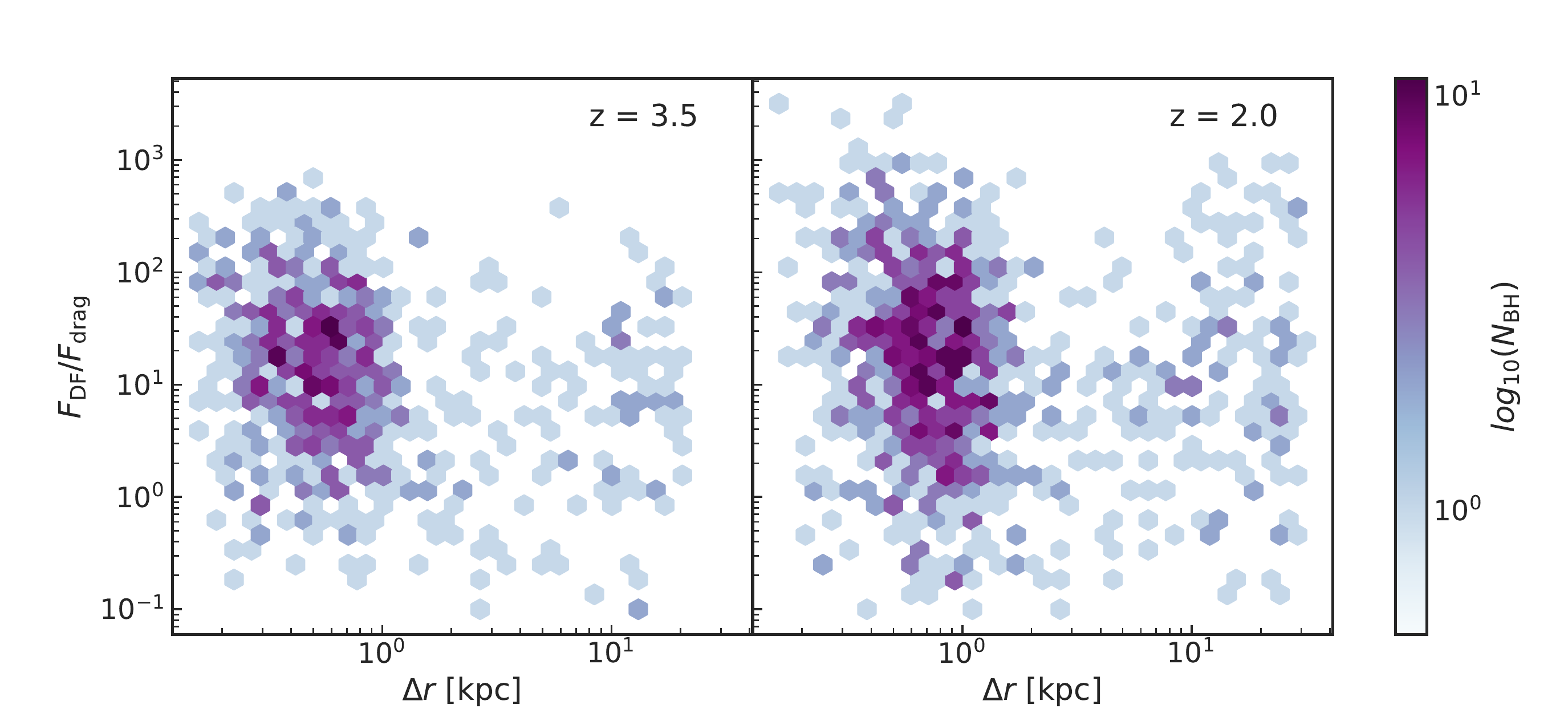}
\caption{\textbf{Left:} Scattering relation between the $F_{\rm DF}/F_{\rm drag}$ ratio and the black hole mass. For each black hole, we sample its mass at uniformly-distributed time bins throughout its evolution, and we show the scattered density of all samples. DF has significantly larger effects over gas drag on larger BHs. We fit the scatter to a power-law shown in the orange line. \textbf{Right:} Scattering relation between the $F_{\rm DF}/F_{\rm drag}$ ratio and the BHs' distance to the halo center. Comparing with the BH mass, we do not see a clear dependence of the $F_{\rm DF}/F_{\rm drag}$ ratio on the distance to halo center. For BHs at all locations within the halo, DF is in general larger than the gas drag.}
\label{fig:drag_scatter}
\end{figure*}

In the previous subsection, we've only included collisionless particles (DM+Star) when modeling the dynamical friction, now we will look into the effects of dynamical friction of gas (gas drag) in comparison with the collisionless particles in the context of our simulations.

From Equation \ref{eq:H14} and \ref{eq:drag}, the relative magnitudes of DF and drag mainly depend on two components: the relative density of DM+stars versus gas, and the values of $\mathcal{F}(x)$ and $\mathcal{I(M)}$. \cite{Ostriker1999} has shown that when a black hole's velocity relative to the medium falls in the transonic regime (i.e. near the local sound speed), $\mathcal{I}$ is a few times higher than $\mathcal{F}$, while in the subsonic and highly supersonic regimes $\mathcal{I}$ is smaller or equal to $\mathcal{F}$. Therefore, we would expect the gas drag to be larger when the black hole is in the early sinking stage with a relatively high velocity and a high gas fraction. 

In Figure \ref{fig:drag}, the left panel shows the comparison between the magnitude of DF and gas drag through different stages of the black hole evolution, as well as the factors that can alter the effectiveness of the gas drag. In the very early stages ($z>7.5$) of black hole evolution, DF and gas drag have comparable effects, while after $z=7.5$ the gas drag becomes significantly less important and almost negligible compared with DF. The reason follows what we have discussed earlier: the gas density decreases relative to the stellar density (shown in the middle panel), and the black hole's velocity relative to the surrounding medium goes into the subsonic regime as a result of the orbital decay (shown in the lower panel). Around $z=3.5$, there is a boost in the black hole's velocity due to disruption during a major merger with a larger galaxy and black hole. The effect of gas is again raised for a short period of time (although still subdominant compared to the DF).

In Figure \ref{fig:big_plot}(b) we plot the black hole evolution for the DF-only (\texttt{DF\_4DM\_G}), drag-only (\texttt{Drag\_4DM\_G}), and DF+drag (\texttt{DF+Drag\_4DM\_G}) simulations.
Both the drag-only and DF-only models are effective in sinking the black hole at early times ($z>7$). However, at lower redshifts, the gas drag is not able to sink the black hole by itself, whereas DF is far more effective in stabilizing the black hole at the halo center. For this reason, in low-resolution cosmological simulations, dynamical friction from collisionless particles is necessary to prevent the drift of the black holes out of the halo center.

To further illustrate the relative importance between DF and gas drag for the entire BH population, we examine the dependencies of the $F_{\rm DF}/F_{\rm drag}$ on variables related to the BH evolution for all BHs in the \texttt{DF+Drag\_4DM\_G} simulation. First, in the right panel of Figure \ref{fig:drag} we show the time evolution of $F_{\rm DF}/F_{\rm drag}$. The top panel shows the ratio as a function of cosmic time, while the bottom panel shows the ratio as a function of each BH's seeding time. The DF/Drag ratio has a wide range for different BHs, but overall DF is becoming larger relative to the gas drag as the black hole evolves. From the mean value of the DF/drag ratio, we see that when the black holes are first seeded, DF is only a few times larger than the gas drag. After a few Gyrs of evolution, DF becomes 2-3 orders of magnitude larger than the gas drag. However, there is not a strong correlation between redshift and the $F_{\rm DF}/F_{\rm drag}$ ratio. 

In the left panel of Figure \ref{fig:drag_scatter}, we show the scattering relation between the $F_{\rm DF}/F_{\rm drag}$ ratio and the black hole mass $M_{\rm BH}$. We see a strong correlation between the $F_{\rm DF}/F_{\rm drag}$ ratio and the black hole mass: DF has significantly larger effects over gas drag on larger BHs, although the range of the ratio is large ar the low mass end. We fit a power-law to the median of the scatter:
\begin{equation}
    \frac{F_{\rm DF}}{F_{\rm drag}} = 250 \left(\frac{M_{\rm BH}}{10^7 M_\odot}\right)^{1.7},
\end{equation}
which roughly characterize the effect of the two forces on BHs of different masses. From this relation we see that for BHs with masses $>10^7 M_\odot$, gas drag is in general less than $1\%$ of DF. Finally, the right panels show the relation between the $F_{\rm DF}/F_{\rm drag}$ ratio and the BH's distance to the halo center: there is not a strong dependency on the BH's position within the halo.


\subsubsection{Comparisons with the T15 Model}
\label{subsec:df100}
\begin{figure}
\includegraphics[width=0.5\textwidth]{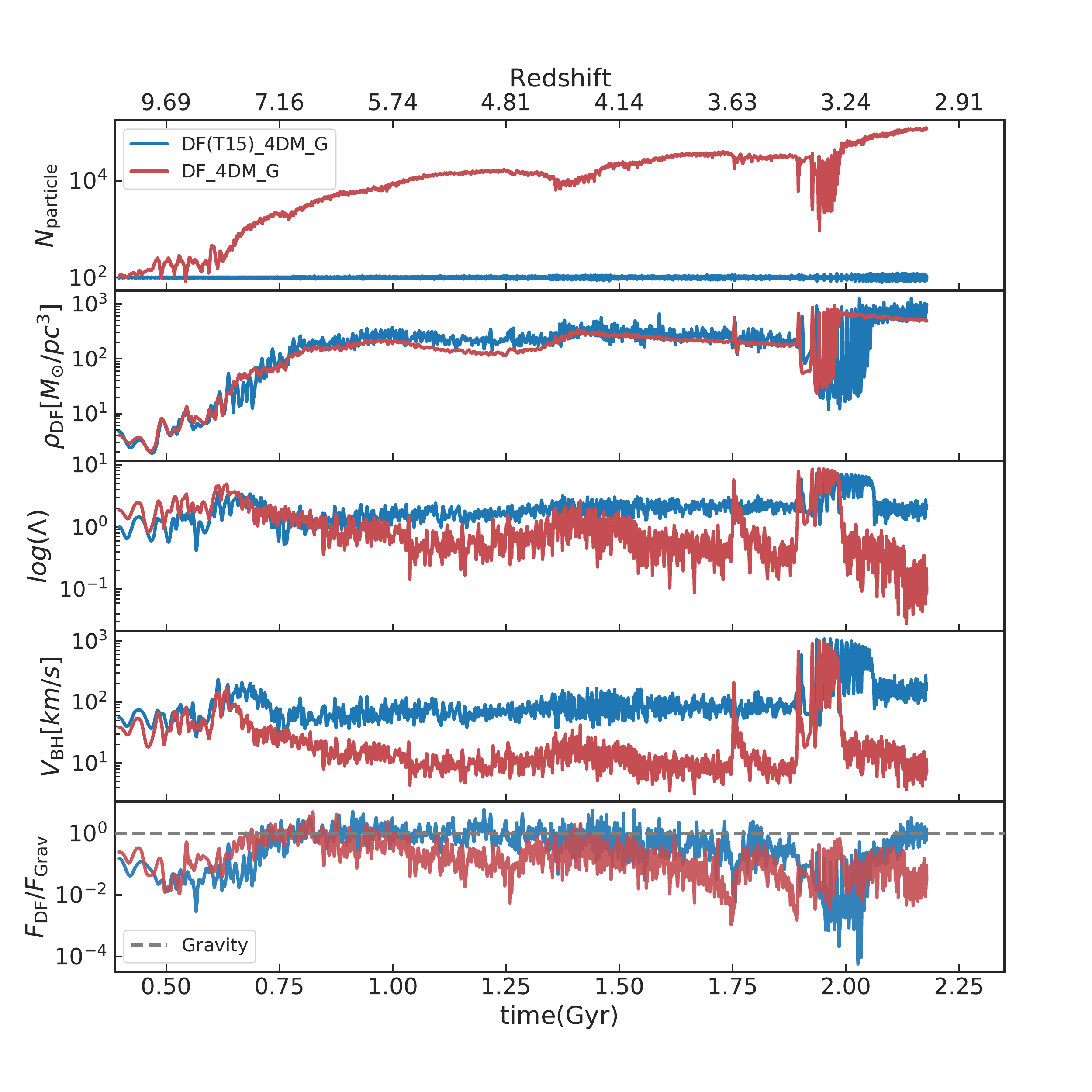}

\caption{ Comparison between different components in the two dynamical friction models, DF(fid) (\textbf{red}) and DF(T15) (\textbf{blue}) (see Section \ref{sec:bh_model} for descriptions). We show the number of stars and dark matter particles included in the DF density and velocity calculation (\textbf{top panel}), the density used for DF calculation (\textbf{second panel}), the Coulomb logarithm used in the two methods (\textbf{third panel}), the velocity of the BH relative to the surrounding particles (\textbf{forth panel}, note that the "surrouding particles" are defined differently for the two models), and the magnitude of DF relative to gravity (\textbf{bottom panel}). The higher DF in the DF(fid) model at $z>8$ is due to the larger Coulomb logarithm. After $z\sim 7$, the higher density of DF(T15) due to more localized density calculation counterbalances its lower $\text{log}(\Lambda)$, resulting in similar DF between $z=8$ and $z=3.5$. During the halo merger at $z=3.5$, the DF(fid) model included particles from the target halo into the density calculation, and therefore yields larger DF during the merger.}
\label{fig:k100_case1}
\end{figure}

For the collisionless particles, we test and study two different implementations for the dynamical friction: DF(fid) and DF(T15) (see Section \ref{sec:bh_model} for detailed descriptions). In Section \ref{sec:bh_model} we pointed out three main differences between them: different kernel sizes (SPH kernel vs. nearest 100 DM+star), different definitions of $b_{\rm max}$ (10 ckpc vs. 1.5 ckpc$/h$), and different approximation of the surrounding velocity distribution (Maxwellian vs. nearest 100-sample distribution). Essentially, these differences mean that DF(fid) is a less-localized implementation than DF(T15). Now we would like to evaluate the effectiveness of these two implementations and show how different factors affect the final dynamical friction calculation.

Figure \ref{fig:k100_case1} shows the relevant quantities in the DF computation for the two methods. The two kernels both contain $\sim 100$ dark matter and star particles at high redshift ($z>8$), but after that the SPH kernel (defined to include the nearest 113 gas particles) begins to include more and more stars and dark matter. The mass fraction of stars in the SPH kernel dominates over that of dark matter by $\sim 10$ times for a BH at the center of the galaxy. The larger kernel of DF(fid) has two effects: first, the DF density will be smoother over time; second, during halo mergers, the DF(fid) kernel can "see" the high-density region of the larger halo, which results in a higher DF near mergers compared to DF(T15). This is confirmed by the second panel, where we show the density for dynamical friction calculation from the two kernels. The densities calculated from the two kernels are similar in magnitude throughout the evolution, although the DF(T15) kernel yields slightly larger density due to its smaller size. Around the BH merger, the density in DF(fid) is larger due to its inclusion of the host halo's central region.

The third panel shows the Coulomb logarithm in the two models. Recall that $\Lambda = \frac{b_{\rm max}}{(GM_{\rm BH})/v_{\rm BH}^2}$, and so the Coulomb logarithm depends on the black hole's mass, its velocity relative to the surrounding particles, and the value of $b_{\rm max}$. From Figure \ref{fig:big_plot}(c), the mass of the DF(T15) black hole is slightly smaller, but the mass difference is small compared with the 6 times difference in $b_{\rm max}$. Given $b_{\rm max}$=10 ckpc$/h$ in DF(fid) and $b_{\rm max}$=1.5 ckpc$/h$ in DF(T15), we would expect the Coulomb logarithm to be larger for the former. However, there is yet another tweak: the $v_{\rm BH}^2$ term turns out to be significantly larger in the DF(T15) model(fourth panel). Note that in the DF(T15) model $v_{\rm BH}^2$ is calculated using only 100 surrounding particles, and for the high-density region we are considering here, the velocity of the nearest 100 particles is very noisy in time.  As we will show in Appendix \ref{app:df100}, for smaller black holes the difference in $v_{\rm BH}^2$ is not as large, and usually DF(fid) has a larger $\text{log}\Lambda$ due to its larger $b_{\rm max}$.

In Figure \ref{fig:big_plot}(c), we show the evolution of the black hole under these two models. At high redshift ($z>8$), due to the large $\text{log}(\Lambda)$, the black hole in the DF(fid) simulation sinks slightly faster to the halo center. Between $z=8$ and $z=3.5$, both models have similar dynamical friction (as discussed in the previous paragraph) and the motion and mass accretion are also similar. Then at $z=3.5$, within the host halo of the black hole major merger, dynamical friction in DF(fid) is again larger because the density kernel includes more particles from the high-density region in the target halo, and this leads to an earlier merger time.

Overall, the performance of the two models is similar. However, as we have seen in the velocity calculation of the black holes relative to the surrounding particles, DF(T15) could be too localized for simulations of our resolution ($\epsilon_g \sim 1$kpc/h) and is sometimes subject to numerical noise. Therefore, in our subsequent statistical runs we pick DF(fid) as our fiducial model, and will drop the 'fid' in its name hereafter.

\subsubsection{Gravitationally Bound Merging Criterion}
\label{subsec:bound_check}

The merging criterion can affect not only the merging time, but also the dynamics and evolution of the black holes. Naively, we might expect the distance-only merging to produce more massive black holes, because black holes are merged more easily. However, in many cases this is not true, and we will illustrate here through one example. 

Figure \ref{fig:big_plot}(d) shows the evolution of the same black hole with the same dynamical friction prescription, but different merging criteria. We note a drastic difference in the black hole's trajectories: while the BH in the gravitationally bound merger case is staying at the center of its host halo, the BH in the distance-only merger flies out of its host after a merger. This is because with the distance-only model, it is possible for one black hole to have a very large velocity at the time of the merger, since we do not limit the black hole's velocity. By momentum conservation, the black hole with a larger velocity can transfer the momentum to the other black hole (and the merger remnant) which might have already sunk to the halo center. The sunk black hole then drifts out of the halo center after a merger due to the large momentum injection. This is especially common in simulations where the black hole's dynamical mass is boosted, because the injected momentum is also boosted with mass and a smaller black hole in a satellite galaxy can easily kick a larger black hole out. If we add on the gravitational bound check, there will be more time for the black holes to lose their angular momentum, and so the injected momentum is far less, and in most cases does not kick each other out of the central region.

\subsection{Black Hole Mergers}
\begin{figure*}
\includegraphics[width=0.49\textwidth]{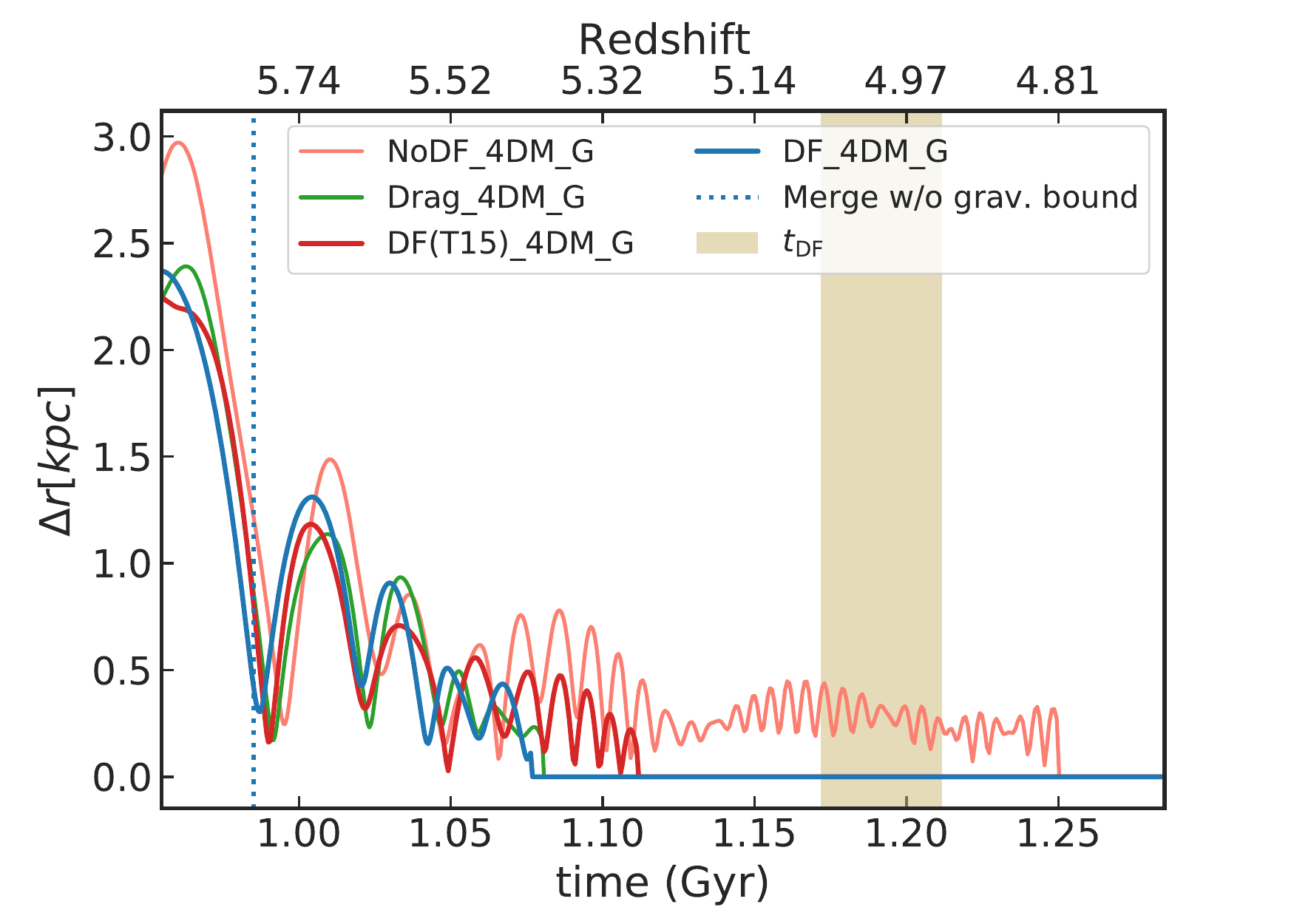}
\includegraphics[width=0.49\textwidth]{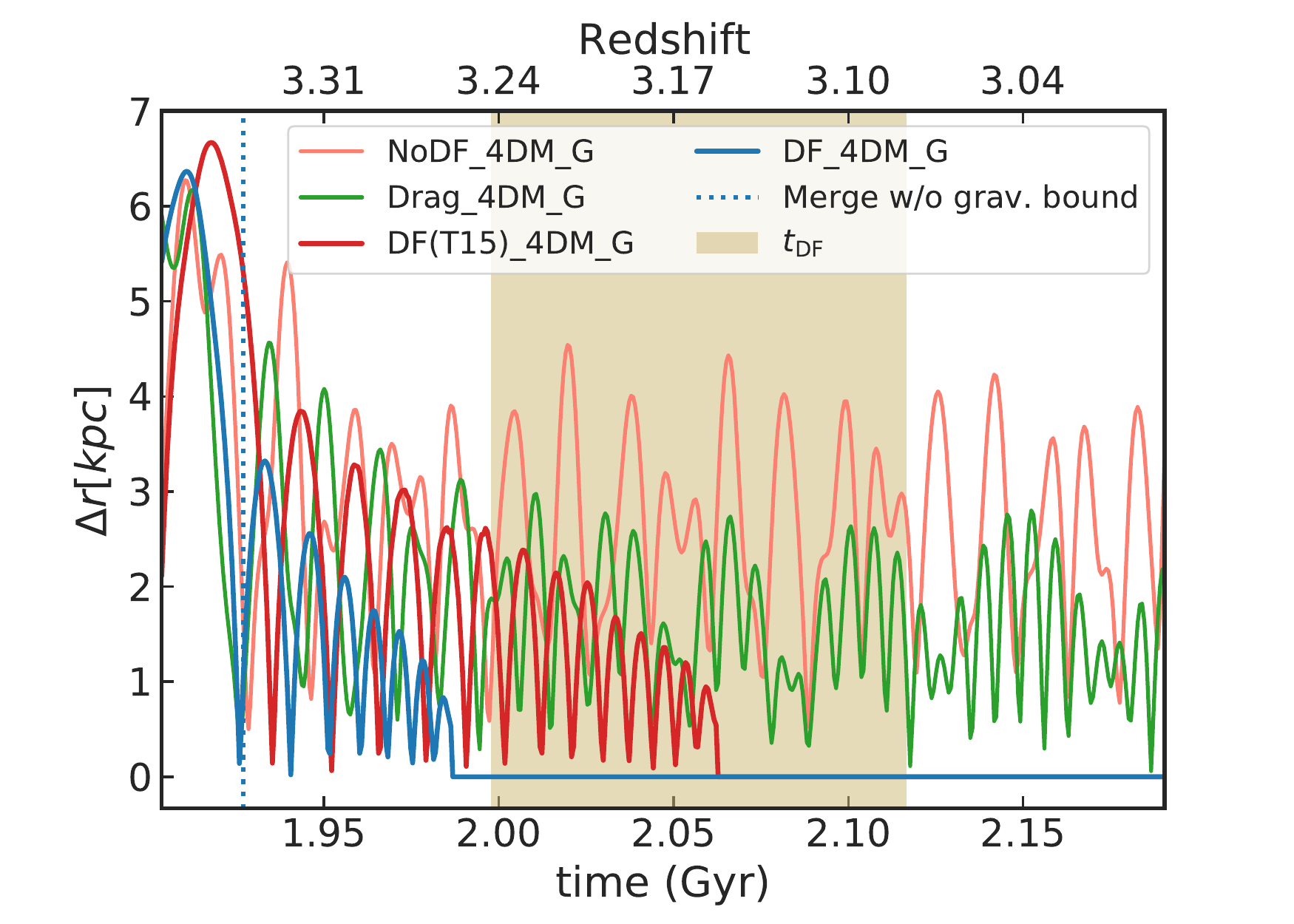}
\includegraphics[width=0.49\textwidth]{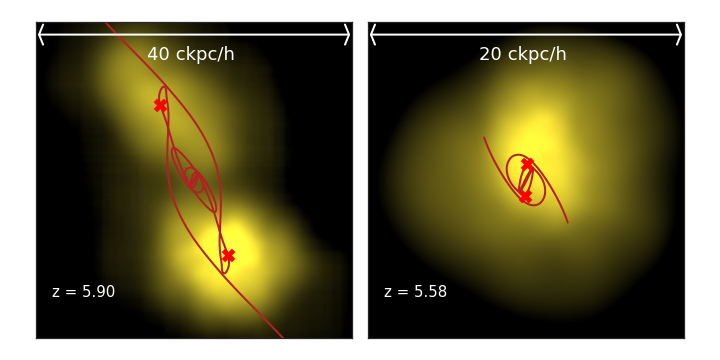}
\includegraphics[width=0.49\textwidth]{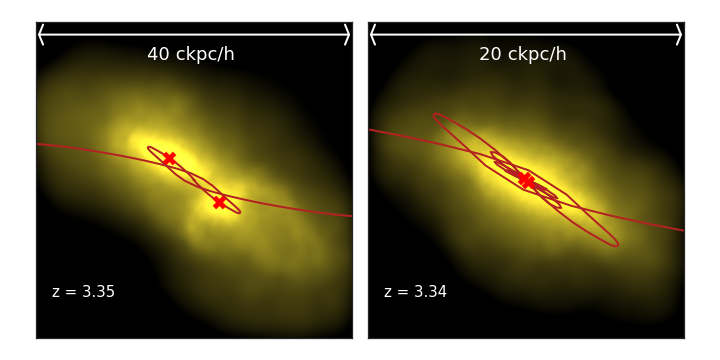}

\caption{The comparison between the distance of two merging black holes in the no-correction, DF(fid), DF(T15) and gas drag models in the early stage (\textbf{left}) and later stage (\textbf{right}) of the black hole evolution. For early mergers, the effect of the frictional forces (DF and drag) is not very prominent but still noticeable. The DF and gas drag both allow the black holes to merge faster compare to the no-DF case. For the later merger happening in a denser environment, the effect of dynamical friction is clear. However, the gas drag does not have a big effect on the black hole at this late stage compared with the no-DF case. The lower panels show the merging black holes within their host galaxies as well as their trajectories towards the merger in the \texttt{DF\_4DM\_G} run. The left images show the early phase of the orbital decay, and the right images show the later phase when the orbits get smaller.}
\label{fig:merger_case1}
\end{figure*}
Having seen the effect of different dynamical models on the evolution of individual black holes, next we will discuss how the dynamics, together with different BH merging criteria,  affect the evolution and mergers of the black holes.
 In particular, we want to study their merging time and trajectories before and after the mergers. Similar to the previous subsection, we will draw our examples from the two halos shown in Figure {\ref{fig:halos}}.
 
\subsubsection{Effect of Dynamical  Friction Modeling}
\label{subsec:case_merger_calc}

We first look at how different dynamical models affect the time scale of black hole orbital decay and mergers. We pick two cases of mergers:  one is an early merger at $z>5$ when the black holes have not outgrown their dynamical masses; the other is a later merger at $z \sim 3.3$ when both BHs are larger than their seed dynamical masses (the major merger between BH1 and BH2 in Figure \ref{fig:halos}). Following \cite{Tremmel2015}, we also compute the dynamical friction time for the two mergers using Equation (12) - Equation (15) from \cite{Taffoni2003}:
\begin{equation}
\label{eq:tdf}
    t_{\rm DF} = 0.6\times 1.67\text{Gyr} \times \frac{r_c^2 V_h}{G M_s} \text{log}^{-1} \left( 1+\frac{M_{\rm vir}}{M_s} \right) \left(\frac{J}{J_c}\right)^\alpha,
\end{equation}
where $M_s$ is the mass of the smaller black hole (which we treat as the satellite), $M_{\rm vir}$ is the virial mass of the host halo of the larger black hole (found by AHF), $V_{h}$ is the circular velocity at the virial radius of the host, and $r_c$ is the radius of a circular orbit with the same energy as the satellite black hole's initial orbit. The last term $\left(\frac{J}{J_c}\right)^\alpha$ is the correction for orbital eccentricity, where $J$ is the angular momentum of the satellite, $J_c$ is the angular momentum of the circular orbit with the same energy as the satellite, and $\alpha$ is given by:
\begin{equation}
    \alpha \left( \frac{r_c}{R_{\rm vir}}, \frac{M_s}{M_{\rm vir}} \right) = 0.475 \left[ 1-\text{tanh} \left( 10.3 \left(\frac{M_s}{M_{\rm vir}}\right)^{0.33} - 7.5 \left(\frac{r_c}{R_{\rm vir}}\right) \right)  \right].
\end{equation}
In our calculation the virial radius, velocity, and mass are obtained from the AHF outputs, and the circular radius, orbit energy, and angular momentum are calculated by fitting the halo density profile to the NFW profile.

Figure \ref{fig:merger_case1} shows distances between two merging black holes in the no-DF, DF(fid), DF(T15), and gas drag models in the early and later stages of their evolution. For the early merger, the effect of the frictional forces (DF and drag) is not very big but still noticeable. The DF and gas drag have similar effects on the orbital decay at higher redshifts, consistent with our discussion in Section \ref{subsec:drag}. The DF(T15) model sinks the black hole a little slower than the DF(fid) model, but the difference is within $50$ Myrs. All three friction models allow the black holes to merge faster compare to the no-DF case by $\sim 150$ Myrs.

For the later merger, which takes place in a denser environment, the effect of dynamical friction is clearer: the dynamical friction allows the black holes to sink within the gravitational softening of the particles in $<200$ Myrs. Without dynamical friction the black hole's orbit does not have a clear decay below $2$ kpc and does not merge at the end of our simulation. Furthermore, the gas drag does not have a big effect on the black hole at this late stage compared with the no-correction case. This follows from our discussion in section \ref{subsec:drag} that gas drag is much less effective at lower redshift compared to dynamical friction.
 
In both plots, the yellow shaded region is the dynamical friction time from the analytical calculation in Equation \ref{eq:tdf}. Here we draw a band instead of a single line, because the black hole's orbit is not a strict ellipse, and the black hole is continuously losing energy. We calculate $t_{\rm DF}$ at multiple points between the first and second peak in the black hole's orbit (e.g. between $z=5.9$ and $z=5.7$ in the earlier case), and plot the range of those $t_{\rm DF}$. For both mergers, the analytical prediction is less than 150 Myrs later than the merger of the (fid) model. We note that the \cite{Taffoni2003} analytical $t_{\rm DF}$ is a fit to the NFW profiles, and the previous numerical and analytical comparisons on the black hole dynamical friction\citep[e.g.][]{Tremmel2015,Pfister2019} are performed in idealized NFW halos with a fixed initial black hole orbit. In our case, the halo profiles and black hole orbits are not directly controlled, and therefore deviation from the analytical prediction is expected. We will study such deviations statistically later in Section \ref{sec:merger_stats}.

 \subsubsection{Effect of Gravitational Bound Check}

 In Section \ref{subsec:merger} we introduced two criteria which we use to perform black hole mergers in our simulations: we can merge two BHs when they are close in distance, and we can also require that the two BHs are gravitationally bounded in addition to the distance check. 
 
 In Figure \ref{fig:merger_case1} we show the difference in black holes' merging time with and without the gravitational bound criterion. The vertical dashed line marks the time that the two black holes in the \texttt{DF\_4DM\_G} simulation would merge if there was not the gravitational bound check. Without the gravitational bound check, the orbit of the black holes is still larger than 1 kpc when they merge, whereas with the gravitational bound check, the orbit size generally decays to less than 300 pc when the black holes merge. The merger without gravitational bound check generally makes the merger happen earlier by a few hundred Myrs (we will study the orbital decay time statistically in the next section). Therefore, for more accurate merger rate predictions as well as the correct accretion and feedback, it is necessary to apply the gravitational bound check during black hole mergers whenever the black hole has a well-defined velocity.

\section{Black Hole Statistics}
\label{sec:stats}
After looking at individual cases of black hole evolution, we now turn to the whole SMBH population in the simulations with different modeling of black hole dynamics. For statistics comparison, instead of using the $L_{\rm box} = 10$ Mpc$/h$ constrained realizations, we now use $L_{\rm box} = 15$ Mpc$/h$ unconstrained simulations. The details of our $L_{\rm box} = 15$ Mpc$/h$ simulations are shown in Table \ref{tab:norm}.

\subsection{Sinking of the Black Holes}
\label{subsec:drift}

\begin{figure}
\includegraphics[width=0.49\textwidth]{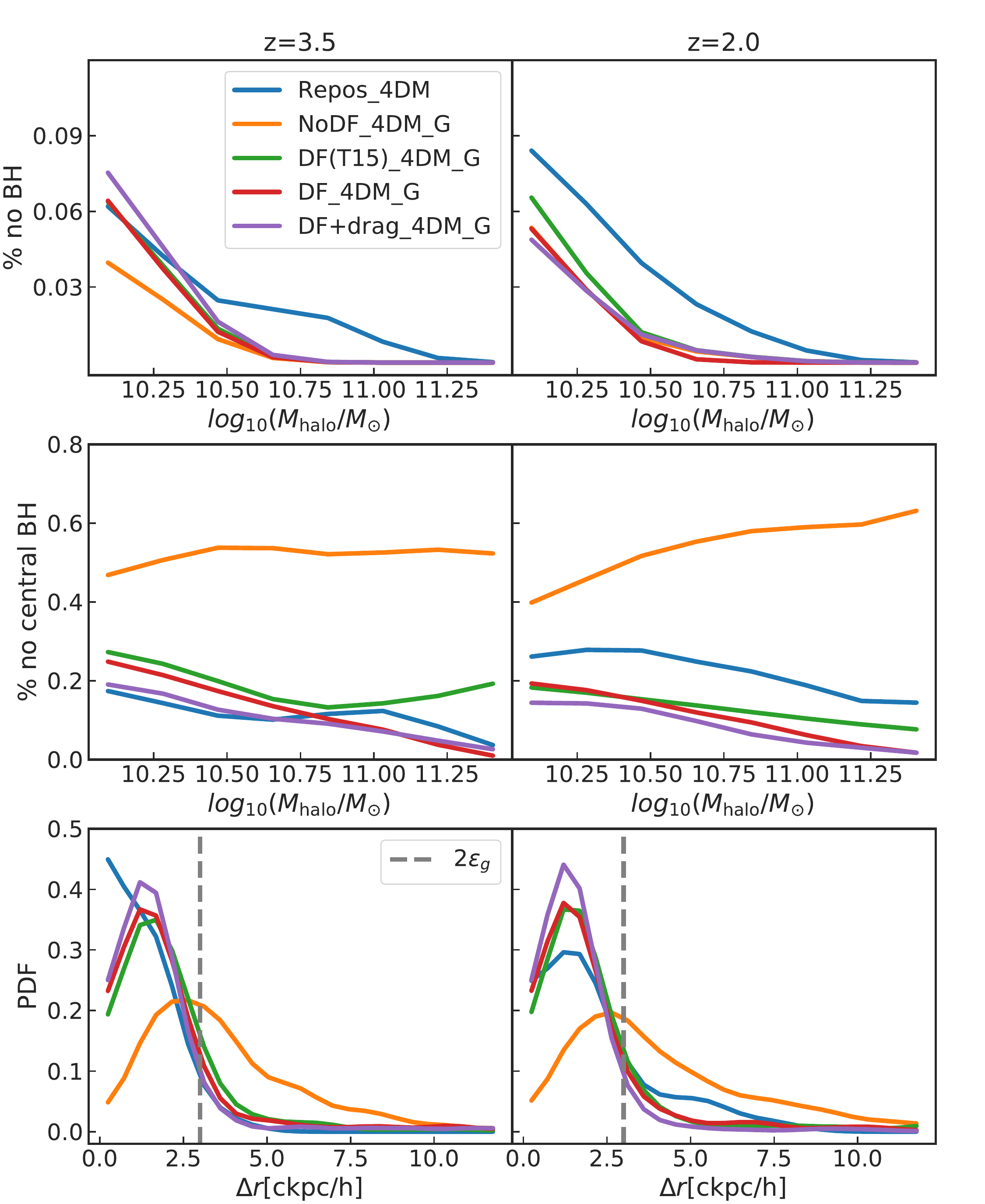}
\caption{The effect of different BH dynamics modeling on BH position relative to its host. We include the reposition model (blue), no-DF model (orange),DF(T15) model (green), DF(fid) model (red) and the DF+drag model (purple). \textbf{Top:} The fraction of halos(subhalos) without a black hole for halos with masses above the black hole seeding mass at $M_{\rm halo} = 10^{10} M_\odot/h$. \textbf{Middle:} The fraction of halos without a central black hole ("central" means within $2\epsilon_g$ from the halo center identified by the halo finder), out of all halos with black holes. \textbf{Bottom:} Distribution of black holes' distance to its host halo center.} 
\label{fig:drift}
\end{figure}

\begin{figure}
\includegraphics[width=0.49\textwidth]{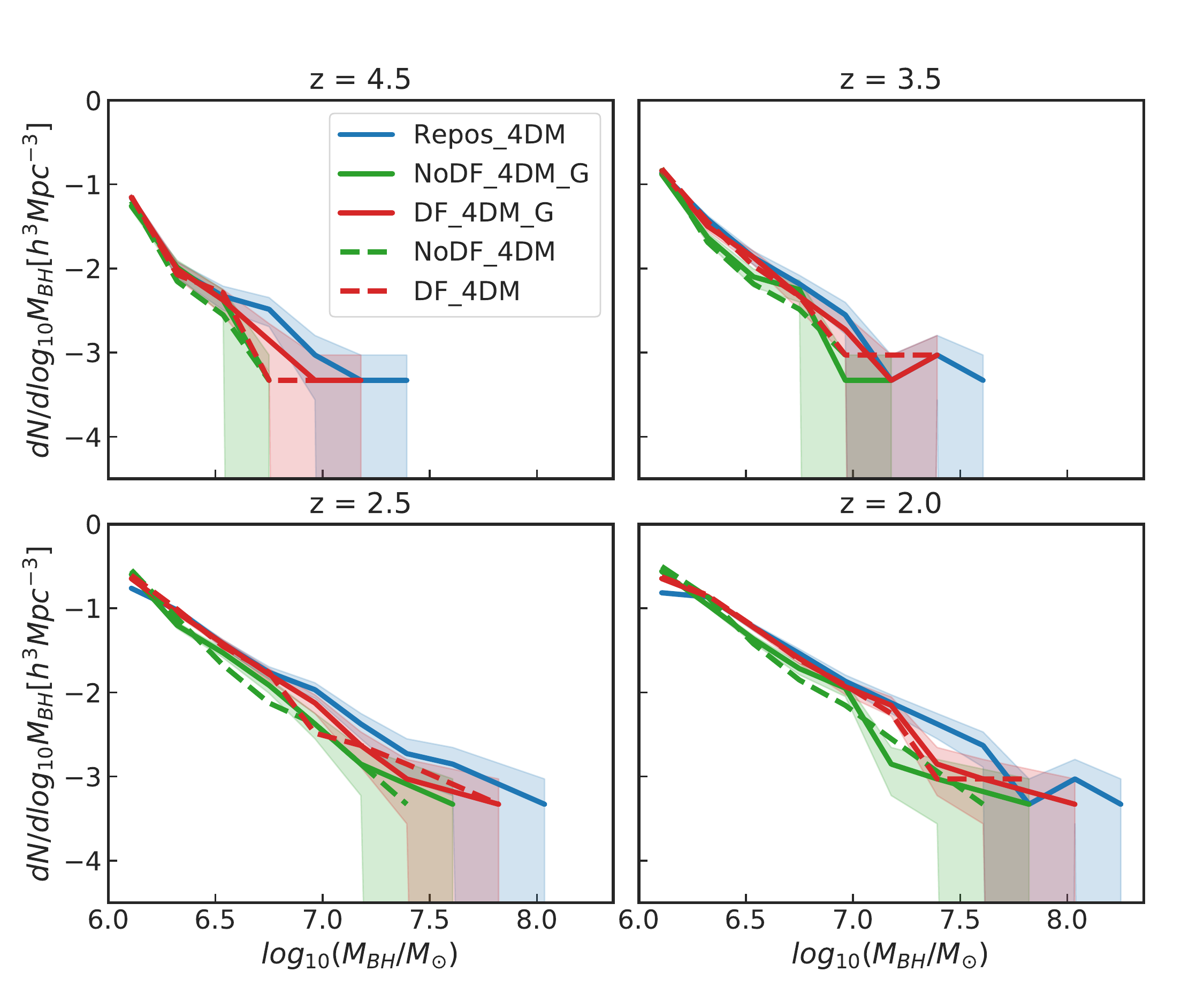}
\caption{Mass functions for reposition, DF and no-DF simulations. With reposition (\textbf{blue}), we have the highest mass function and earlier formation of $10^8 M_\odot$ black holes. The no-DF simulations (\textbf{green}) have lower mass functions, which is expected due to low-accretion and merger rates from the black hole drifting. The dynamical friction model (\textbf{red}) yields a mass function in between.} 
\label{fig:bhmf}
\end{figure}

The primary reason for adding dynamical friction onto black holes within the cosmological simulations is to stabilize the black hole at the halo center (defined as the position of the minimum-position gas particle within the halo). Hence, we start by looking at the black holes' position relative to the host halos. Due to the resolution limit of our simulations, we would not expect the black holes to be able to sink to the exact minimum potential. Instead we consider a $<2\epsilon_g = 3$ ckpc$/h$ distance to be "good sinking".

In Figure \ref{fig:drift}, we show the statistics related to black holes' sinking status. We included the comparison between the reposition model (\texttt{L15\_Repos\_4DM}), the no-DF model (\texttt{L15\_NoDF\_4DM}), the two dynamical friction models (\texttt{L15\_DF\_4DM} and \texttt{L15\_DF(T15)\_4DM}) and the DF+drag model(\texttt{L15\_DF+drag\_4DM}). To start with, we simply count the fraction of halos without a black hole when its mass is already above the black hole seeding criterion (i.e. $10^{10}M_{\odot}/h$). The top panel shows the fraction of large halos without a BH for different models at $z=3.5$ and $z=2$. Surprisingly, the no-DF model ends up with the least halos without a black hole. This is because even though the black holes without dynamical corrections cannot sink effectively, the high dynamical mass still prevents sudden momentum injections from surrounding particles, and therefore most BHs still stays within their host galaxies. The dynamical friction models perform equally well, with $<10\%$ no-BH halos at the low-mass end. The reposition model, however, ends up with the most no-BH halos, even though repositioning is meant to pin the black holes to the halo center. This happens because under the repositioning model, the central black holes tend to spuriously merge into a larger halo during fly-by encounters, leaving the smaller sub-halo BH-less.

Next we look at where the black holes are located within their host galaxies. For all the halos with at least one black hole, we examine whether the black hole is located at the center (i.e.$<2\epsilon_g = 3$ckpc/h from the halo center). The middle panel of Figure \ref{fig:drift} shows the fraction of halos without a central BH. The no-DF model has significantly more halos without a central BH compared to the other models, with over half of the halos hosting off-center BHs. Among the three runs with dynamical friction, the DF(T15) and DF(fid) models have a similar fraction of halos ($\sim 20\%$) without a central BH, and we can see this fraction dropping from $z=3.5$ to $z=2$, meaning that many BHs are still in the process of sinking towards the halo center. When we further add the gas drag, $10\%$ more halos host at least one central BH, and the difference between the drag and no-drag central BHs is more prominent at high redshifts. 

Interestingly, the repositioning algorithm is not as efficient at sinking the BHs at $z=2$ as the DF. This is because our repositioning algorithm places the BHs at the minimum potential position within the accretion kernel, instead of within the entire halo. The majority of the offset between the BH positions and the halo center comes from the offset between the minimum-potential position accessible to the BH (i.e. minimum-potential in the accretion kernel) and the minimum-potential position in the halo. Such offset can be especially severe at lower redshift, when the size of the accretion kernel gets smaller and mergers happen more frequently, making it easier for the black holes to get stuck at a local minimum.

In the bottom panels we show the distributions of the black holes' distance to the halo centers under different models. For the no-DF run, again we see that the black holes fail to move towards the halo center at lower redshift, resulting in a much flatter distribution compared to all the other models. In comparison, when we add dynamical friction to the black holes, for both the DF(fid) and the DF(T15) models the distributions are pushed much closer to the halo center, with a peak around the gravitational softening length. When we then add the gas drag in addition to DF, the peak at $\epsilon_g$ becomes slightly higher than those in the DF-only runs. The combination of DF and gas drag, as we would expect from the case studies, is the most effective in sinking the black holes to the halo centers and stabilizing them. Finally, we plot the repositioning model for reference.  It does well in putting the black hole close to the minimum potential, and often the black holes can be located at the exact minimum-potential position (the distributions peak at 0 for $z=3.5$). However, as discussed in the previous paragraph, there are cases where the local minimum potential found by the repositioning algorithm does not coincide with the global minimum potential of the halo, and that is why we also see non-zero probability density for $\Delta r > 3$ ckpc$/h$ at $z=2$.

The statistics we have seen for the models above are consistent with the results from the case studies. This shows that even though for the case studies we have focused mainly on large black holes in one of the biggest halo, a similar trend still applies to other black holes in the cosmological simulations, which are embedded in smaller halos or subhalos.


\subsection{Black Hole Mass Function}
\label{subsec:bhmf}
Next we look at how different dynamics affect the black hole mass function (BHMF). One problem with the repositioning method is that it places the black holes at the galaxy center too quickly, which could result in excess accretion and thus a higher mass function. On the other hand, if we do not add any correction to the black hole motion, many BHs will not go though efficient accretion and mergers, and we will see a lower mass function. We would expect the BHMF in the dynamical friction run to fall between the repositioning case and the no-DF case.

Figure \ref{fig:bhmf} shows the BHMF from the reposition(\texttt{L15\_Repos\_4DM}), dynamical friction (without gravitational bound check:\texttt{L15\_DF\_4DM}; with gravitational bound check:\texttt{L15\_DF\_4DM\_G}), and no-DF (without gravitational bound check: \texttt{L15\_NoDF\_4DM}; with gravitational bound check: \texttt{L15\_NoDF\_4DM\_G}) runs. The reposition model yields the highest mass function, and is the only simulation with more than one $10^8 M_{\sun}/h$ black holes at $z=2$. This is expected from the over-efficient BH mergers and the high-density surroundings in the reposition model. Moreover, it creates increasingly more massive BHs over time, as the increased merger rate produces a stronger effect over time. The no-DF runs produces the lowest mass function due to the off-centering, while the DF mass function falls between the reposition and no-DF case as we expected. 

Naively, we would expect the models without gravitational bound checks to produce a higher mass function, because it allows for easier mass-accretion via mergers. However, as discussed in Section \ref{subsec:bound_check}, this is not the case if we compare the dashed lines and solid lines with the same colors. For example, under the DF model, the \texttt{L15\_DF\_4DM\_G} simulation forms more massive black holes than the \texttt{L15\_DF\_4DM} simulation, especially at lower redshift. The reason can be traced back to what we have seen in Figure \ref{fig:big_plot}(d): when there is no gravitational bound check, the large momentum injection during a merger kicks the black hole out of the halo center, thus preventing the efficient growth of large black holes.

Considering the relatively large uncertainties due to the limited volume, the difference in the mass function is not very significant. We would expect other factors such as the black hole seeding, accretion and feedback to have a larger effect on the mass function compared to the dynamical models we show here \citep[e.g.][]{Booth2009}.
\subsection{Dynamical Friction Time and Mergers}
\label{sec:merger_stats}

\begin{figure}
\includegraphics[width=0.48\textwidth]{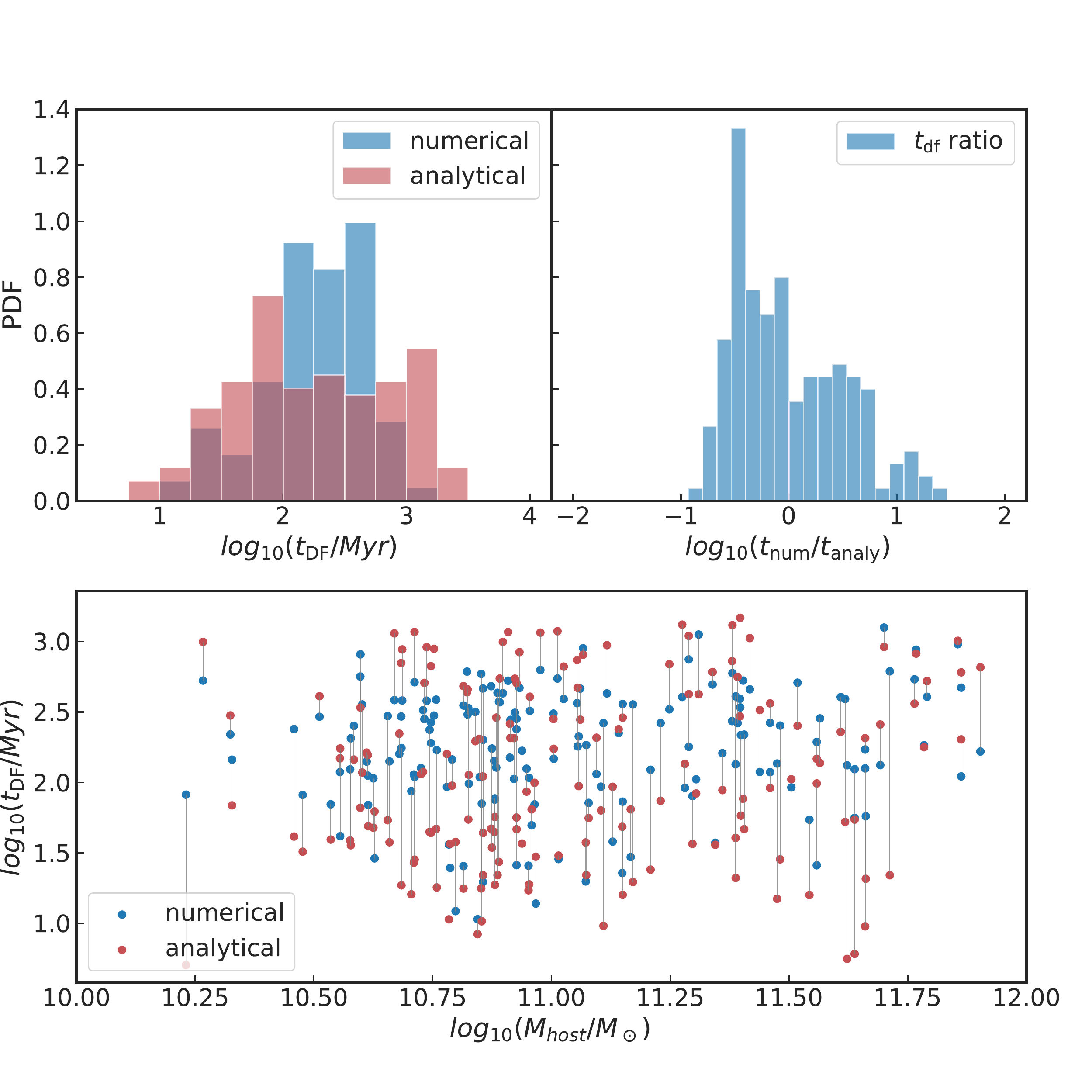}
\caption{The delay of mergers due to the dynamical friction time. Here we compare the numerical dynamical friction time,$t_{\rm num}$, to the analytically calculated time (following Equation \ref{eq:tdf}) $t_{\rm analy}$. \textbf{Top left:} distribution of the dynamical friction time from numerical merger (blue) and analytical predictions (red). \textbf{Top right:} ratio between the numerical and analytical $t_{\rm df}$. Their difference is less than one order of magnitude in all merger cases. \textbf{Bottom:} dynamical friction time as a function of the virial mass of the host halo for the numerical (blue) merger and analytical predictions (red). The same merger event is linked by a grey line.} 
\label{fig:delay}
\end{figure}

\begin{figure}
\includegraphics[width=0.49\textwidth]{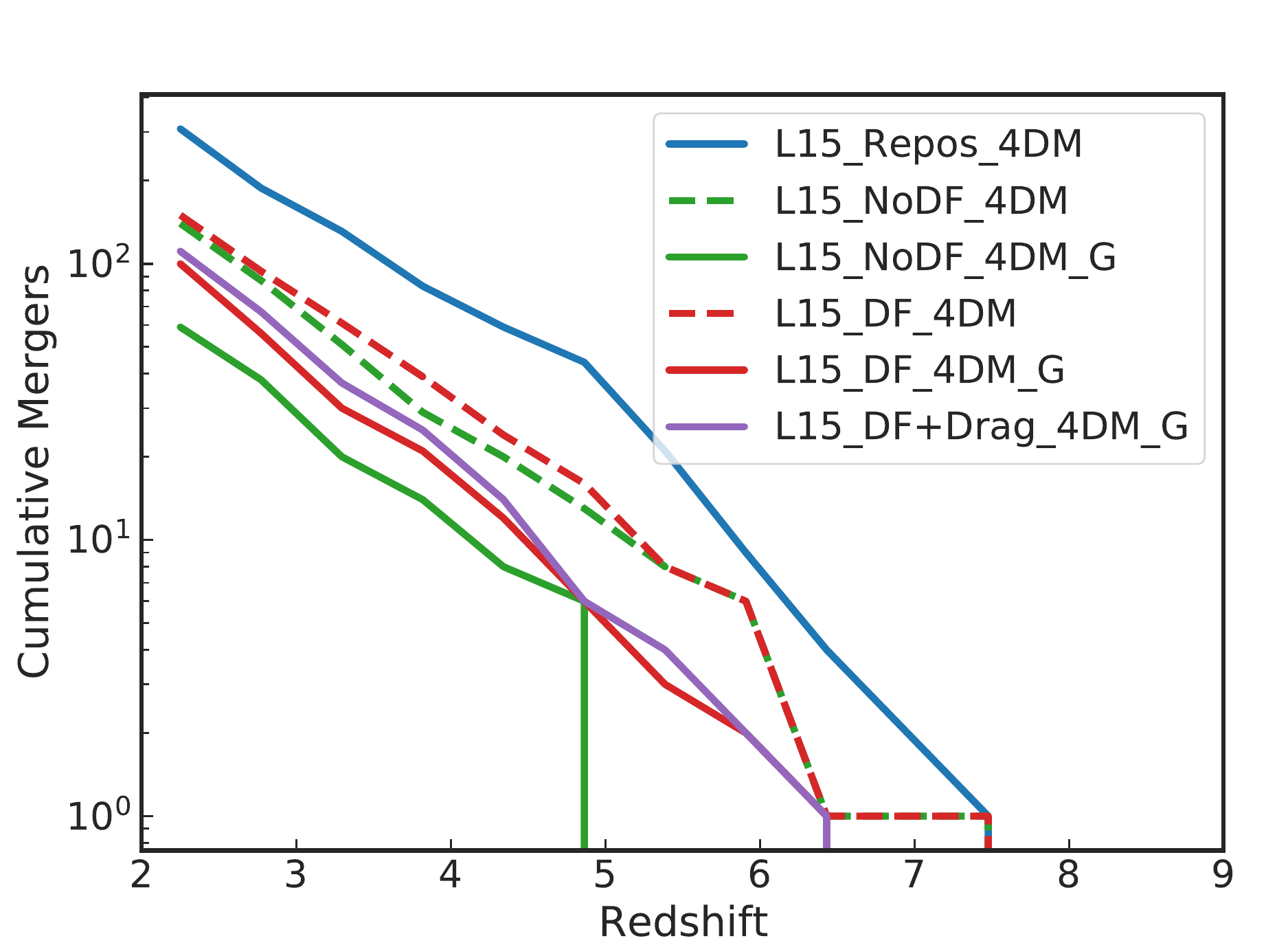}
\caption{The cumulative mergers for different BH dynamics and merging models. The reposition model (\textbf{blue solid}) predicts more than two times the total mergers compared with the other models. Without the gravitational bound check, the DF (\textbf{red dashed}) and the no-DF model (\textbf{green dashed}) predicts similar numbers of mergers, indicating that the first encounters of the black hole pairs are similar under the two models. However, if we add the gravitational bound check, the dynamical friction model (\textbf{red solid}) yields $\sim 50\%$ more mergers compared to the no-correction model. Adding the gas drag in addition to dynamical friction (\textbf{purple solid}) raises the mergers by a few. } 
\label{fig:merger_stats}
\end{figure}

Because the reposition method is used in most large-volume cosmological simulations, a post-processing analytical dynamical friction time is calculated in order to make more accurate merger rate predictions. Now that we have accounted for the dynamical friction on-the-fly, we want to study how our numerical mergers with dynamical friction compare against the analytical predictions, and how different dynamical models impact the black hole merger rate.

In Section \ref{subsec:case_merger_calc}, we compared the numerical merging time to the analytical predictions for two merger cases. Now we use the same method to calculate an analytical dynamical friction time for all black hole mergers in our \texttt{L15\_DF\_4DM\_G} simulation. For each pair, we begin the calculation at the time $t_{\rm beg}$ when the black hole pair first comes within 3 ckpc$/h$ of each other, as this mimics the merging time without the gravitational bound check, and is also close to the merging criterion under the reposition model. The numerical dynamical friction time $t_{\rm num}$ is the time between the numerical merger and $t_{\rm beg}$. The analytical dynamical friction time $t_{\rm analy}$ is calculated using the host halo information in the snapshot just before $t_{\rm beg}$ and the black hole information at the exact time-step of $t_{\rm beg}$.

Figure \ref{fig:delay} shows the comparison between the numerical and analytical dynamical friction times. In the top panel we show the distribution of the two times as well as the distribution of their ratio. We note that for all the mergers happening numerically, $t_{\rm analy}$ does not exceed 2 Gyrs, and most have $t_{\rm analy}$ less than 1 Gyr. This means that we do not have many fake mergers that shouldn't merge until much later (or never). Also, the ratio plot shows that the numerical and analytical times are always within an order of magnitude of each other, with most of the numerical mergers earlier than the analytical mergers. The numerical merger time is peaked between 100 Myrs and 1 Gyrs, whereas the analytical calculation yields a flatter distribution. We would expect $t_{\rm analy}$ to be longer than $t_{\rm num}$, both because we have a selection bias on $t_{\rm DF}$ by ending the similation at $z=2$, and because we numerically merge the black holes when their orbit is still larger than 3 ckpc$/h$. However, this does not explain why $t_{\rm analy}$ has a higher probability between 10 Myrs and 100 Myrs. 

To see the individual merger cases in the distribution more clearly, in the lower panel of Figure \ref{fig:delay} we plot all the numerical and analytical dynamical friction times as a function of the host halo's virial mass. From this figure we do not see a clear dependence of either dynamical friction times on the host halo's virial mass. There is also no strong correlation between the $t_{\rm num}/t_{\rm analy}$ ratio and the halo mass. We do not further investigate the discrepancies between the numerical and analytical results, as these results can vary significantly from system to system. 

We note that although the numerical model has free parameters (such as $b_{\rm max}$, $M_{\rm dyn, seed}$) that can impact the merging time (but see Appendix \ref{app:merger_param}), it can account for the immediate environment around black hole and adjust the dynamical friction on-the-fly. More importantly, it also accounts for the interaction between the satellite BH and its own host galaxy, which could reduce the sinking time significantly \citep[e.g.][]{Dosopoulou2017}. 
The analytical model, though verified by N-body simulations, does not react to the environment of the merging galaxies by always assuming an NFW profile. Moreover, it only models the sinking of a single BH without embedding it in its host galaxy. Therefore, we expect the numerical result to be a more realistic modeling of the binary sinking process.

After comparing the DF model against the analytical prediction, next we compare different numerical models in terms of the black hole merger rate. Figure \ref{fig:merger_stats} shows the cumulative mergers from $z=8$ to $z=2$. We have included comparisons between the reposition, dynamical friction and no-DF models, both with and without the gravitational bound check. The reposition model predicts more than twice the total number of mergers compared to the other models. Without the gravitational bound check, the DF and the no-DF models predict similar numbers of mergers, indicating that the first encounters of the black hole pairs are similar under the two models. However, if we add the gravitational bound check, the DF model yields $\sim 50\%$ more mergers compared to the no-DF model, because the addition of DF assists energy loss of the binaries and leads to earlier bound pairs. Finally, the merger rate is not very sensitive to adding the gas drag: the merger rate in the DF-only model is similar to that of the DF+drag model. This can be foreseen in the comparison shown in Figure \ref{fig:drag}, where the gas drag is subdominant in magnitude.

\section{Merger Rates in the 35Mpc/h Simulations}
\label{sec:L35}

\begin{figure*}
\includegraphics[width=0.99\textwidth]{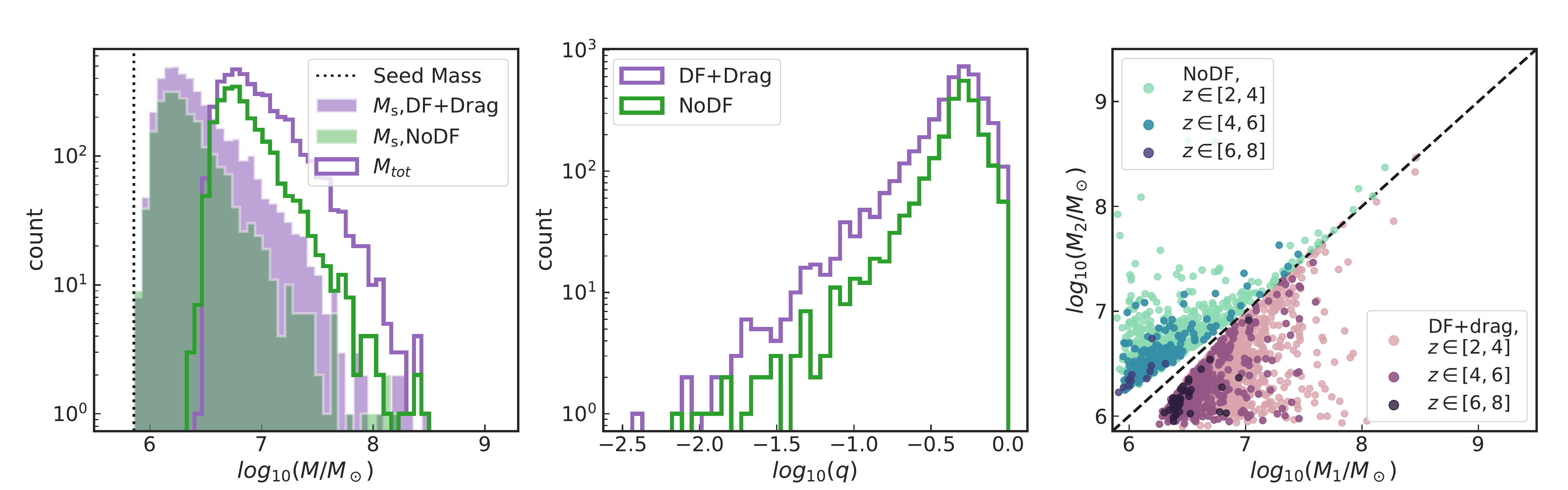}
\caption{ \textbf{Left:} Distribution of the mass of the smaller black hole ($M_s$), and distribution of the total mass of the binary ($M_{\mathrm{tot}}$). For both simulations, the mergers in which at least one of the black holes is slightly above the seed mass dominate. The most massive binary has a total mass of $3\times 10^8 M_\odot$. \textbf{Middle:} The mass ratio $q$ between the two black holes in the binary. We see a peak at $\text{log(q)}=-0.5$, corresponding to pairs in which one BH is about three times larger than the other. \textbf{Right:} Scatter of the two black hole masses in the binaries, binned by redshift. To separate the scatter in the two simulations, for the DF+drag run we take $M_1$ to be the mass of the larger BH, while for the NoDF run $M_2$ is the larger BH.}
\label{fig:hist}
\end{figure*}

\begin{figure}
\includegraphics[width=0.49\textwidth]{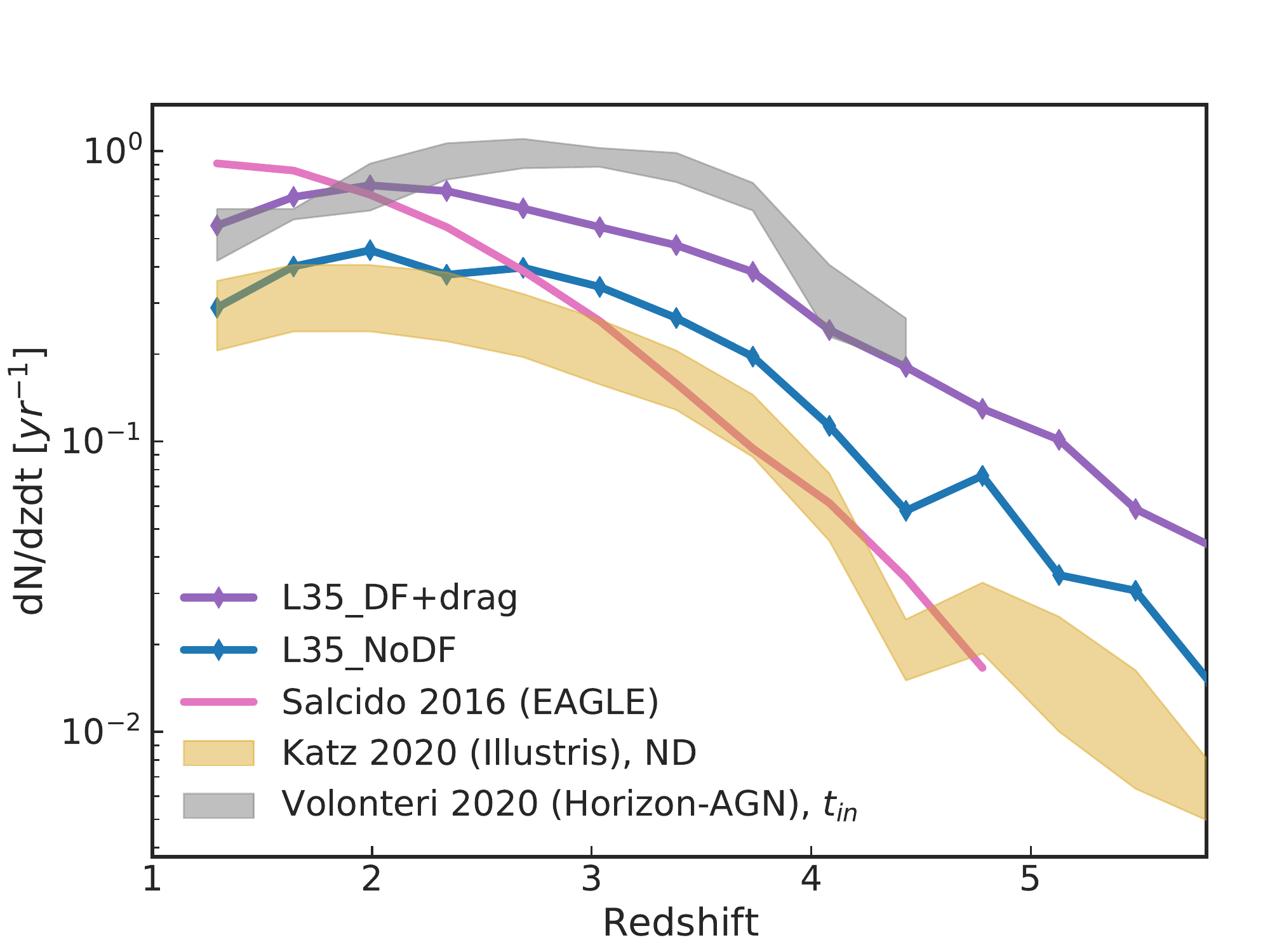}
\caption{Merger rate per year of observation per unit redshift predicted from our \texttt{L35\_DF+drag\_4DM\_G} (\textbf{purple}) and \texttt{L35\_NoDF\_4DM\_G} (\textbf{blue}) simulations. 
For comparison, we also show the the prediction from recent hydro-dynamical simulations. 
We include three simulations of similar mass-resolution: \citet{Volonteri2020} from the Horizon-AGN simulation (\textbf{gray}), \citet{Katz2020} (\textbf{yellow}) from the Illustris simulation and \citet{Salcido2016} from the EAGLE simulations (\textbf{pink}).
Since we do not apply any post-processing delays after the numerical mergers, we only compare to results without delays.}
\label{fig:rates}
\end{figure}

Based on all the previous test of BH dynamics modeling, we have reached the conclusion that the DF+drag model with $M_{\rm dyn} = 4 M_{\rm DM}$ is most capable of sinking the black hole to the halo center. Hence, we choose to use this model to run our larger-volume simulation \texttt{L35\_DF+drag\_4DM\_G} for the prediction of the BH coalescence rate. Besides this model, we also perform a same-size run without the dynamical friction, \texttt{L35\_NoDF\_4DM\_G}, as a lower limit for the predicted rate.  
Our \texttt{L35} simulations are run down to $z=1.1$. The black hole seed mass is $5\times10^5 M_\odot/h$ and the minimum halo mass for seeding is $10^{10} M_\odot/h$. The details of these two simulations are shown in Table \ref{tab:norm}.

\subsection{The Binary Population}
\label{subsec:L35_catalog}
Because this work mainly focuses on model verification and is not intended for accurate merger-rate predictions, we do not account for the various post-numerical-merger time delays. These delays can be caused by physical processes such as sub-ckpc scale dynamical friction, scattering with stars, gravitational wave driven inspiral and triple MBH systems \citep[e.g.][]{Quinlan1996,Sesana2007b,Vasiliev2015,Dosopoulou2017,Bonetti2018}. We consider all the numerical mergers as true black hole merger events. Without any post-process selection, there are 25224 black holes and 4237 mergers in the \texttt{L35\_DF+drag\_4DM\_G} run, and 27693 black holes and 2349 mergers in the \texttt{L35\_NoDF\_4DM\_G} run down to $z=1.1$.

Figure \ref{fig:hist} shows the distribution of the binary parameters for the mergers in our simulations. For both simulations, there is at least one black hole around the seed mass for most mergers, but the peak does not lie at the exact seed mass. The most massive binary has a total mass of $3\times 10^8 M_\odot$. For the mass ratio $q$ between the two black holes in the binary, we see a peak at $\text{log}(q)=-0.3$, corresponding to pairs in which one BH is about two times larger than the other. Finally, we show the scatter of the two progenitor masses. The low mass end of the population deviates more from $q=1$, while the majority of same-mass mergers come from the $5\times 10^6 M_\odot\sim 5\times 10^7 M_\odot$ mass range.

Comparing with previous simulations such as \cite{Salcido2016,Katz2020}, we do not see as many cases of seed-seed mergers, but our distribution in q is similar to that shown in \cite{Weinberger2017} where the larger progenitor is a few times larger than the small progenitor. This is due to our larger black hole seed mass of $5\times10^5 M_\odot$ ($10^6 M_\odot$ in \cite{Weinberger2017}): the mass accretion in the early stage is proportional to $M_{\rm BH}^2$, and so during the time before the black hole mergers, our black holes accrete more mass compared to the simulations with smaller seeds. This explains why both of our black holes in the binaries are not peaked at the exact seed mass.

\subsection{Merger Rate Predictions}
\label{subsec:L35_rates}
We use the binary population shown in the previous section to predict the merger rate observed per year per unit redshift.
The merger rate per unit redshift per year is calculated as:
\begin{equation}
     \frac{dN}{dz\;dt} =  \frac{N(z)}{\Delta z V_{c,sim}} \frac{dz}{dt} \frac{dV_c(z)}{dz}\frac{1}{1+z},
\end{equation}
where $N(z)$ is the total number of mergers in the redshift bin $z$, $\Delta z$ is the width of the redshift bin, $V_{c,sim}$ is the comoving volume of our simulation box and $dV_c(z)$ is the comoving volume of the spherical shell corresponding to the $z$ bin. 

We compare our results against recent predictions from hydro-dynamical simulations of similar resolution, \cite{Salcido2016}, \cite{Katz2020} and \cite{Volonteri2020}. Here we briefly summarize relevant information about their merger catalogs. The Ref-L100N1504 simulation in the EAGLE suite used in \cite{Salcido2016} has an $2^3$ times larger simulation box and slightly higher resolution than our simulations. They seed $1.4\times 10^5M_\odot$ black holes in $1.4\times 10^{10}M_\odot$ halos. They adopt the reposition algorithm for black hole dynamics, but set a distance and relative speed upper limit on the repositioning to prevent black holes from jumping to satellites during fly-by encounters. We compare with their no-delay rate during the inspiral phase. The \textit{Illustris} simulation used in \cite{Katz2020} has a similar box size, resolution and BH dynamics to the Ref-L100N1504 simulation in EAGLE, except that their halo mass threshold for seeding BHs is $7\times 10^{10} M_\odot$. We compare against their ND model, in which mergers are also taken to occur at the numerical merger time without any delay processes. The Horizon-AGN simulation in \cite{Volonteri2020} is $4^3$ times larger than our simulation box, with $\sim 5$ times coarser mass resolution and a black hole seed mass of $10^5 M_\odot$. Instead of seeding BHs in halos above certain mass threshold, the seeding in \cite{Volonteri2020} is based on the local gas density and velocity dispersion, and seeding is stopped at $z=1.5$. For black hole dynamics, they apply dynamical friction from gas, but not from collisionless particles.

Figure \ref{fig:rates} shows our merger rate prediction in the \texttt{L35\_DF+drag\_4DM\_G} and  \texttt{L35\_NoDF\_4DM\_G} simulations. The \texttt{L35\_DF+drag\_4DM\_G} run predicts $\sim 2$ mergers per year of observation down to $z=1.1$, while the \texttt{L35\_NoDF\_4DM\_G} run predicts $\sim 1$. The merger rates from both simulations peak at $z\sim 2$. This factor-of-two difference between the two simulations is consistent with what we predicted in the $L_{\rm box} = 15$ Mpc$/h$ runs in Figure \ref{fig:merger_stats}. Although we did not run a $L_{\rm box} = 35$ Mpc$/h$ simulation with the repositioning model, we expect such a run to predict $5\sim 6$ mergers per year down to $z=1.1$ according to \ref{fig:merger_stats}.

Generally speaking, our simulations yield similar merger rates as the raw predictions from the previous works of comparable resolution. However, we still note some differences both in the overall rates and in the peak of the rates. We will now elaborate on the reasons for those discrepancies.

First, both of our simulations predict more mergers compared with the \cite{Katz2020} ND model prediction. This is surprising given that in the 15 Mpc$/h$ runs we saw $2\sim 3$ times more mergers when we used the reposition method like \cite{Katz2020} and \cite{Salcido2016} did, comparing to our DF+Drag model. Although \cite{Katz2020} cut out $\sim 25\%$ secondary seed mergers and binaries with extreme density profiles, their rate is still lower after adding the cut-out population. One major reason for the higher rate from our simulation compared to \cite{Katz2020} is the different seeding parameters we use: our minimum halo mass for seeding a black hole is $10^{10} M_\odot/h$, which is 5 times smaller compared with \cite{Katz2020}. Moreover, our seeds are a factor of 5 larger. Hence, we have a denser population of black holes in less-massive galaxies, which is likely to result in a higher merger rate even compared to the reposition model used in \textit{Illustris}.

Second, although the rates from EAGLE, Horozon-AGN and our \texttt{L35\_DF+drag\_4DM\_G} simulation cross over at $z\sim 2$, the slope of our merger rate is very different. \cite{Volonteri2020} predicts most mergers at $z\sim 3$, whereas the \cite{Salcido2016} rate peaks at $z\sim 1$. This difference can also be traced to the different seeding rate in the three simulations: in \cite{Salcido2016}, the seeding rate keeps increasing until $z\sim 0.1$, while we observe a drop in seeding rate at $z=3$ in our simulations. In \cite{Volonteri2020}, due to the different seeding mechanism, BH seeds form significantly earlier, leading to a peak in merger rate at a higher redshift. Hence the peak in the BH merger rate is strongly correlated with the peak in the BH seeding rate.

Finally, besides the effect due to different BH seed models on the merger rate, higher resolution can significantly increase the BH merger rates in the simulations. As was shown in previous work \citep[e.g.][]{Volonteri2020,Barausse2020}, dwarf galaxies in low-mass halos can have large numbers of (small mass) BH mergers, and so resolving such halos and galaxies can increase the BH merger rate significantly. The merger rate differences between high and low resolution and the associate choice for the seed models can lead to large differences in the predictions of merger rates than taking account DF in the BH dynamics. 

\section{Conclusions}
\label{sec:conclusion}
In this work we have tested methods for implementing dynamical friction from collisionless particles (i.e. stars and dark matter) and gas in low-resolution cosmological simulations (with mass resolution $M_{\rm DM} \sim 10^7 M_{\odot}$ and spatial resolution of $\epsilon_g \sim$1kpc/h), both for single black hole evolution/mergers using constrained simulations, and for the black hole population using unconstrained simulations.

We showed that dynamical friction from collisionless particles can effectively assist the black hole orbit to decay to within $2\epsilon_g$ of the galaxy center, representing a marked improvement over models that do not include any dynamical correction. Importantly, we find that for our prescription to work well, the dynamical mass of the black holes must be at least twice the mass of the dark matter particles. This is in agreement with results from \cite{Tremmel2015}. The dynamical friction implementation from \cite{Tremmel2015} (DF(T15)) and our implementation adapted to lower-resolution simulations (DF(fid)) result in dynamical friction of a similar magnitude, and have comparable effects on the black holes' dynamics. However, we find that our fiducial model is marginally more suitable for low-resolution simulations, as the nature of the calculation results in less noisy force corrections.

After applying the dynamical friction and performing the gravitational bound check on the black hole pairs, the dynamical friction time of the black holes is consistent with analytical predictions, although the variances can be large for individual black holes due to their varied environments. We note that checking whether the two black holes are gravitationally bound at the time of the merger is necessary both for preventing sudden momentum injection onto the black holes, and for allowing a more realistic orbital decay time.

By direct comparison of the force magnitudes throughout the simulation, we find that dynamical friction from collisionless particles dominate in the majority of cases. The influence of gas drag is highest at the high redshifts, but even then it is typically similar to or less than the contribution from stars and dark matter. This is in broad agreement with the results from \cite{Pfister2019}, though we stress that our simulations cannot resolve the structure of gas on the smallest scales. It is possible that interactions with gas is still important, such as migration within circumbinary disks \citep[e.g.][]{Haiman2009}.

Using our fiducial DF+drag model, we calculate the cumulative merger rate down to $z=1.1$ using a $L_{\rm box}=35$ Mpc$/h$ simulation. Without considering any post-merger delays, we predict $\sim 2$ mergers per year for $z>1.1$, and we lower bound our prediction by a no-dynamical-friction run which predicts $\sim 1$ merger per year. 
Compared with existing predictions from hydro-dynamical simulations \citep[][]{Salcido2016,Katz2020,Volonteri2020}, our rates are consistent with the raw merger rates (rates before post-processing delays are added) from previous works of similar resolution. 
While the dynamics modeling has significant effects (factor of a few according to our experiments) on the BH merger rate, we also found that the different BH seeding criteria and mechanisms account also play a big role in the merger rate predictions.

 Our work has demonstrated the feasibility of recovering sub-kpc-scale BH dynamics in low-resolution cosmological simulations by adding the unresolved dynamical friction. This is the first step in improving upon the widely-adopted reposition model and in tracking the BH dynamics directly down to the resolution limit. Beyond the resolution limit, we still need to account for several smaller-scale binary processes before we can make realistic merger rate predictions \citep[e.g.][]{Quinlan1996,Sesana2007b,Haiman2009,Vasiliev2015,Dosopoulou2017,Bonetti2018,Kelley2017,Katz2020}. Nevertheless, having access to the full dynamical information of the binary at the time of the numerical merger also helps us to better model these small-scale processes. We will leave the analysis of post-merger delays for future works.

There are still several aspects of the DF model that remain somewhat uncertain. Most importantly, the parameters (e.g. $b_{\rm max}$,$M_{\rm dyn,seed}$) in the current dynamical friction model can induce uncertainties in the sinking timescale and the merger rate predictions. For example, reducing $M_{\rm dyn,seed}$ to a value similar to or below the dark matter particle mass will reduce the merger rate by a factor of two or more. Our current choice is well tested in our simulations, but it is still subject to the limitations of our spatial and mass resolution. The limit in the $M_{\rm BH}/M_{\rm DM}$ ratio also hinders comprehensive studies of BH seeding scenarios in the cosmological context. We would need insights from high-resolution simulations \citep[e.g.][]{Dosopoulou2017,Pfister2019} to better model the dynamics of low-mass BHs within cosmological simulations.

\section*{Acknowledgements}
We thank Marta Volonteri for discussions on the merger rate comparisons. The simulations were performed on the Bridges and Vera clusters at the Pittsburgh Super-computing Center (PSC). TDM acknowledges funding from NSF ACI-1614853, NSF AST-1616168, NASA ATP 19-ATP19-0084 NASA ATP 80NSSC18K101, NASA ATP NNX17AK56G, and 80NSSC20K0519.
SPB was supported by NSF grant AST-1817256.


\bibliographystyle{mnras}
\bibliography{main}



\appendix

\section{Dynamical Mass and Resolution Effect}
\label{app:res}
\begin{figure*}
\includegraphics[width=0.9\textwidth]{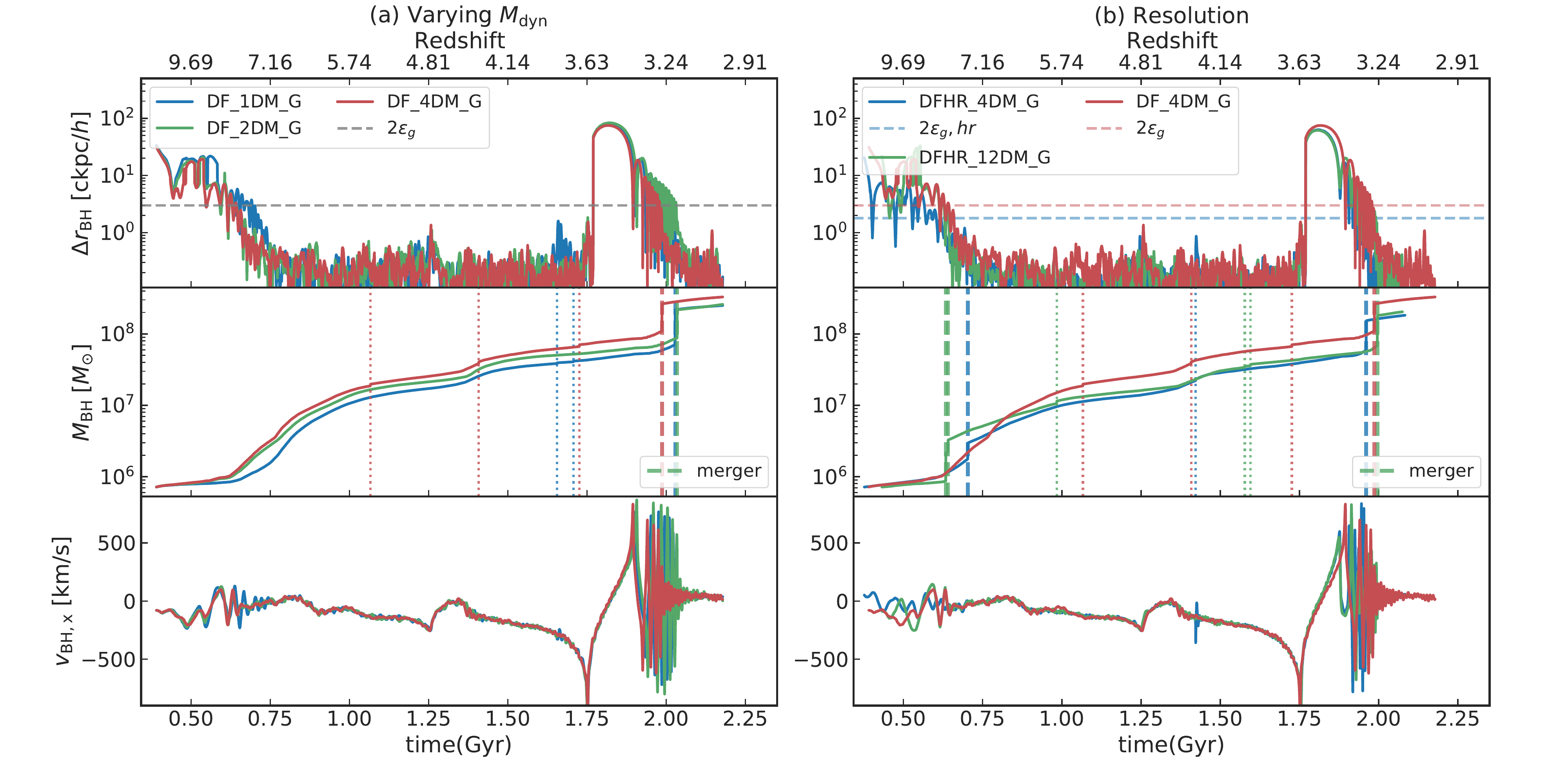}

\caption{\textbf{(a)}: Comparisons of different black hole seed dynamical mass. The effect of varying $M_{\rm dyn,seed}$ is small in this case. But this is partially due to the large BH we pick. \textbf{(b):} Comparison with higher-resolution run with the same $M_{\rm dyn}/M_{\rm DM}$ ratio.}
\label{fig:app_res}
\end{figure*}

\subsection{Varying Dynamical Mass}

One major difference between our model and previous modeling of the dynamical friction is that we boost the mass term during the early stage of black hole growth by a factor of $k_{\rm dyn} = M_{\rm dyn,seed}/M_{\rm BH}$. This is to prevent the drifting of the black holes due to dynamical heating when the black hole mass is below the dark matter particle mass in the context of large and low-resolution cosmological simulations.

Here we show the effect of setting different $k_{\rm dyn}$ by running three simulations with the same resolution and dynamical friction models, but various $k_{\rm dyn}$ ratios. They are listed in Table \ref{tab:cons} as \texttt{DF\_4DM\_G}, \texttt{DF\_2DM\_G}, and \texttt{DF\_1DM\_G}, with $k_{\rm dyn}=4,2,1$, respectively.

Figure \ref{fig:app_res}(a) shows the evolution of the same black hole for different $k_{\rm dyn}$. By comparing the three cases, we can see that the black hole's behavior is very similar for all the physical quantities we have plotted. However, we also note that the similar behavior of different $M_{\rm dyn}$ is case-dependent. The case we present here is a black hole within a large density peak where the black hole is subject to a deep potential and can sink more easily, but the sinking of BHs in shallower potentials can be more sensitive to the seed dynamical mass. Nevertheless, $k_{\rm df}=2$ is generally sufficient to assist the sinking of most black holes and produces similar merger rates to $k_{\rm df}=4$ (see Appendix \ref{app:merger_param}). The convergence at $k_{\rm df}=2$ is consistent with the $M_{\rm BH}/M_{\rm DM}=3$ ratio used in \cite{Tremmel2018}, and relaxes the ratio used in previous works \citep[e.g.][]{Tremmel2015,Pfister2019} of $M_{\rm BH}\sim 10M_{\rm DM}$.

\subsection{Resolution Effect}

Here we show how our model performs under different resolutions. For this experiment we use our fiducial resolution run \texttt{DF\_4DM\_G}, a higher resolution run  \texttt{DF\_HR\_4DM\_G} with the same $k_{\rm df}$, but a factor of three difference in the mass resolution, and a high resolution run \texttt{DF\_HR\_12DM\_G} with the same $M_{\rm dyn,seed}$ as the fiducial resolution run. We would want the black holes to behave similarly independent of resolution if the $M_{\rm dyn,seed}/M_{\rm DM}$ is kept constant.

Figure \ref{fig:app_res}(b) shows the same black hole in the simulations with different resolution. In the high-resolution run \texttt{DF\_HR\_4DM\_G}, even though the seeding dynamical mass is 3 times smaller than the low-resolution run, the sinking time remains the same. Furthermore, if we keep the absolute seeding dynamical mass the same in the low-resolution and high-resolution runs (by comparing \texttt{DF\_HR\_12DM\_G} with \texttt{DF\_4DM\_G}), the black holes still shows similar evolution. This indicates that a constant $k_{\rm df}  = M_{\rm dyn,seed}/M_{\rm DM}$ is robust under different resolutions, and our model of dynamical mass does converges to the true black hole mass if we go to higher resolutions.


\section{DF(fid) vs. DF(T15): cases of smaller black holes evolution}
\label{app:df100}
\begin{figure*}
\includegraphics[width=0.33\textwidth]{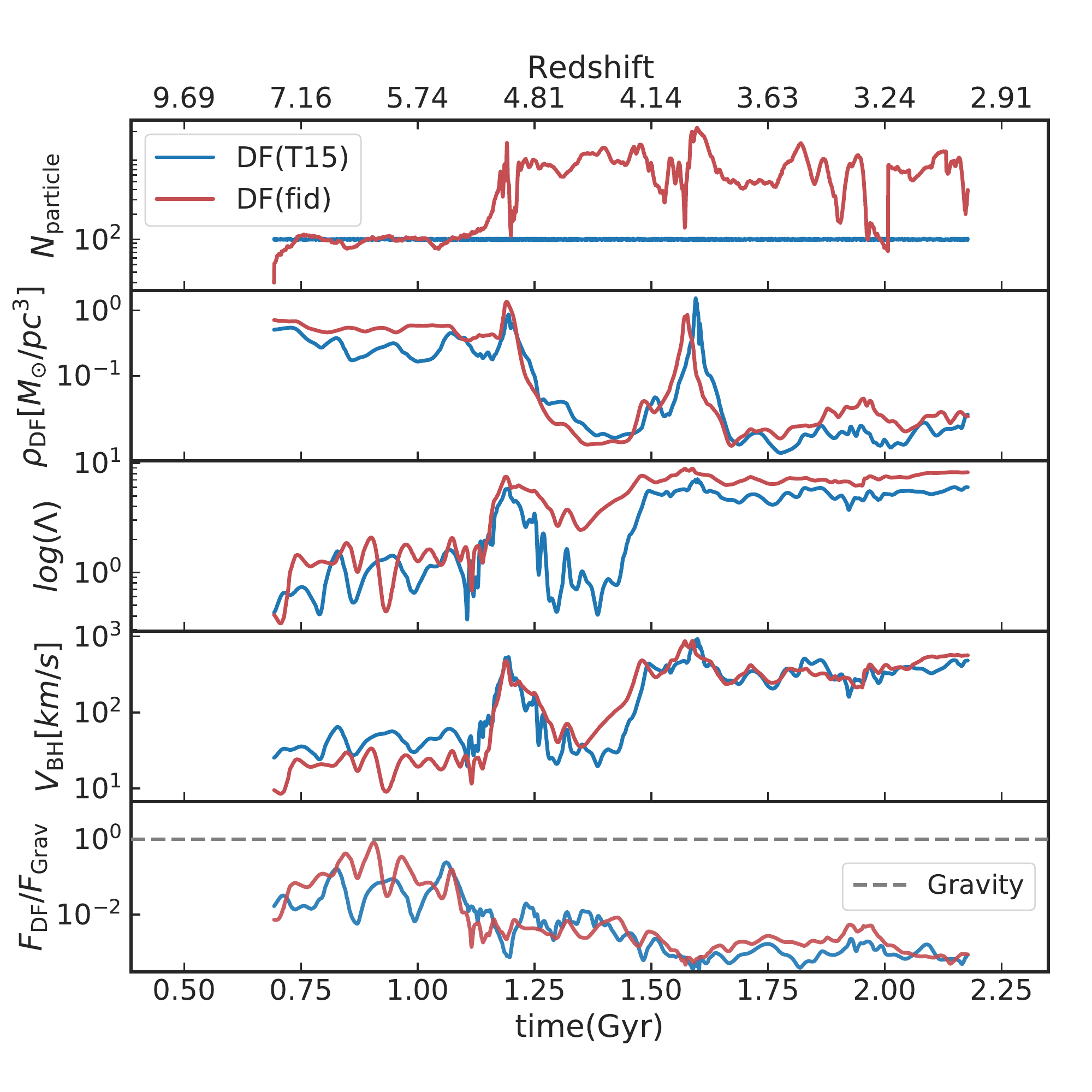}
\includegraphics[width=0.33\textwidth]{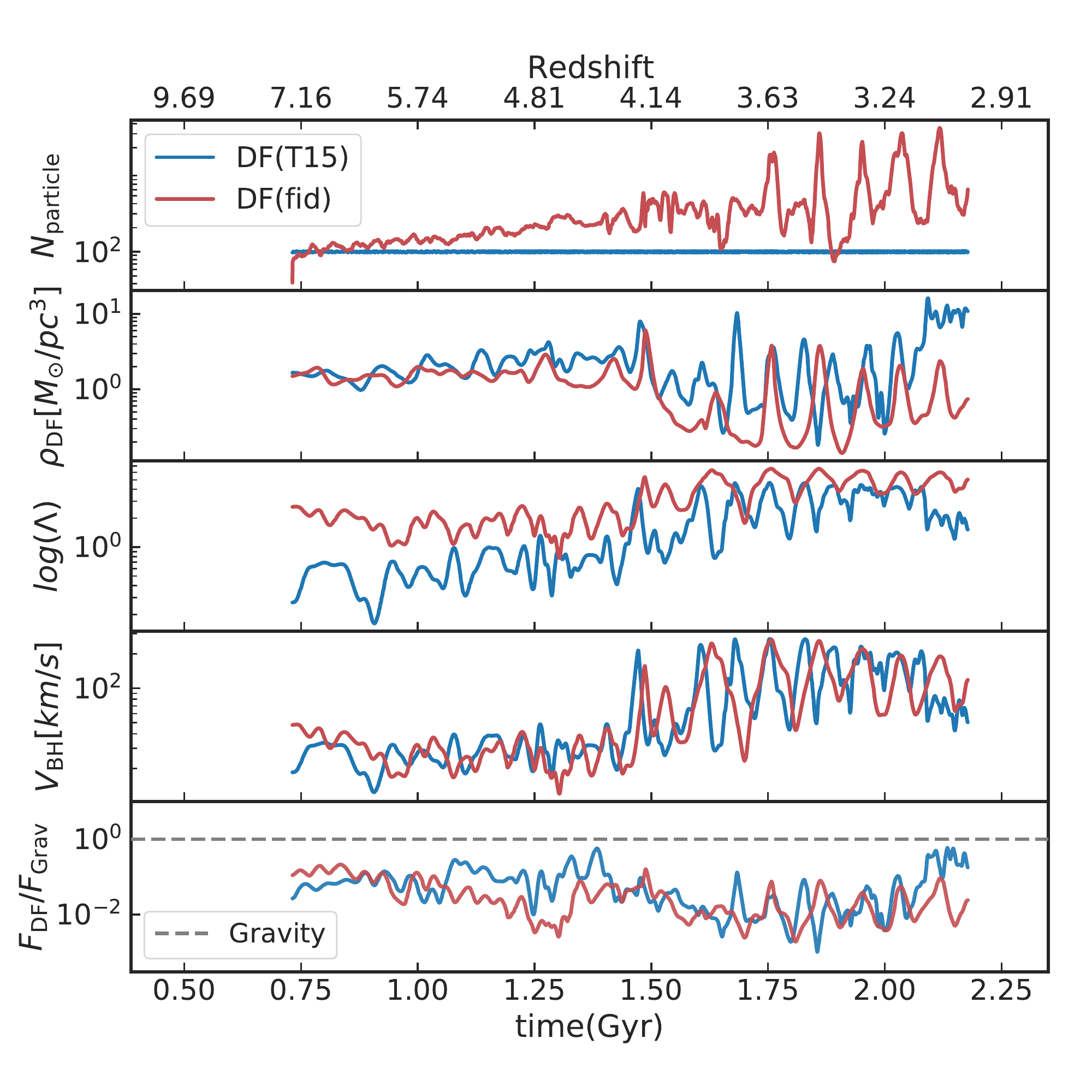}
\includegraphics[width=0.33\textwidth]{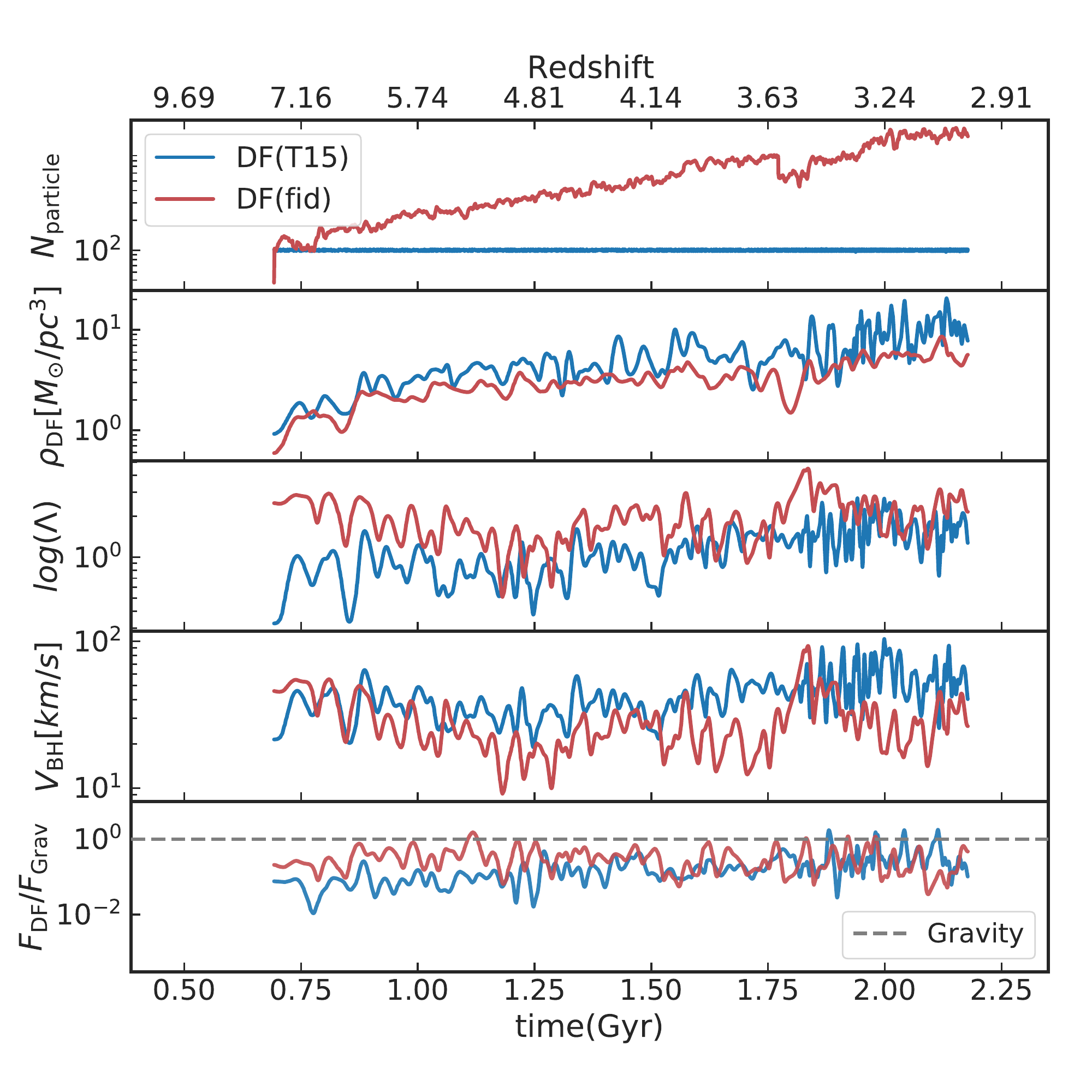}
\caption{Components of the dynamical friction in the \texttt{DF(fid)\_4DM\_G} (\textbf{red}) and the \texttt{DF(T15)\_4DM\_G} (\textbf{blue}) simulations, for three $M<5\times 10^6 M_\odot$ black holes. In these cases, the number of particles within the SPH kernel is still at least an order of magnitude more than 100 at lower redshift. The value of the Coulomb logarithm is now mainly affected by $b_{\rm max}$, because we do not see as much noise in the velocity of the surrounding particles as in the case of a very large BH. In all three cases shown, the magnitude of the dynamical friction is similar in the two models.}
\label{fig:app_kernel}
\end{figure*}

In Section \ref{subsec:models}, we compared the two DF models by showing the example of an early forming black hole located at the center of the largest halo in the simulation. However, that black hole might not be representative of the entire BH population due to its early seeding and large mass. Now we pick more cases of smaller black holes to demonstrate the differences/similarities between the models. In particular, we will look at how the smaller BHs are affected by the DF(fid)/DF(T15) implementation.

Figure \ref{fig:app_kernel} shows the evolution of three small BHs in the \texttt{DF\_4DM\_G} and the \texttt{DF(T15)\_4DM\_G} simulations. We plot three $M_{\rm BH}<5\times 10^6 M_\odot$ black holes. In these cases, the number of particles within the SPH kernel is still at least an order of magnitude more than 100 at lower redshift, and so the density calculated in DF(T15) still tends to be larger but more noisy. The value of the Coulomb logarithm is now mainly affected by $b_{\rm max}$, because we do not see as much noise in the velocity of the surrounding particles as in the case of a very large BH. The density and the Coulomb logarithm counteract each other, and the magnitude of the dynamical friction is similar in the two models. 

These cases again verifies that the two models are consistent with each other, with DF(T15) a more localized implementation than DF(fid). The choice of DF(fid) as our fiducial model is mainly due to our resolution limit.

\section{Effect of Model Parameters on the Merger Rate}
\label{app:merger_param}
\begin{figure}
\includegraphics[width=0.5\textwidth]{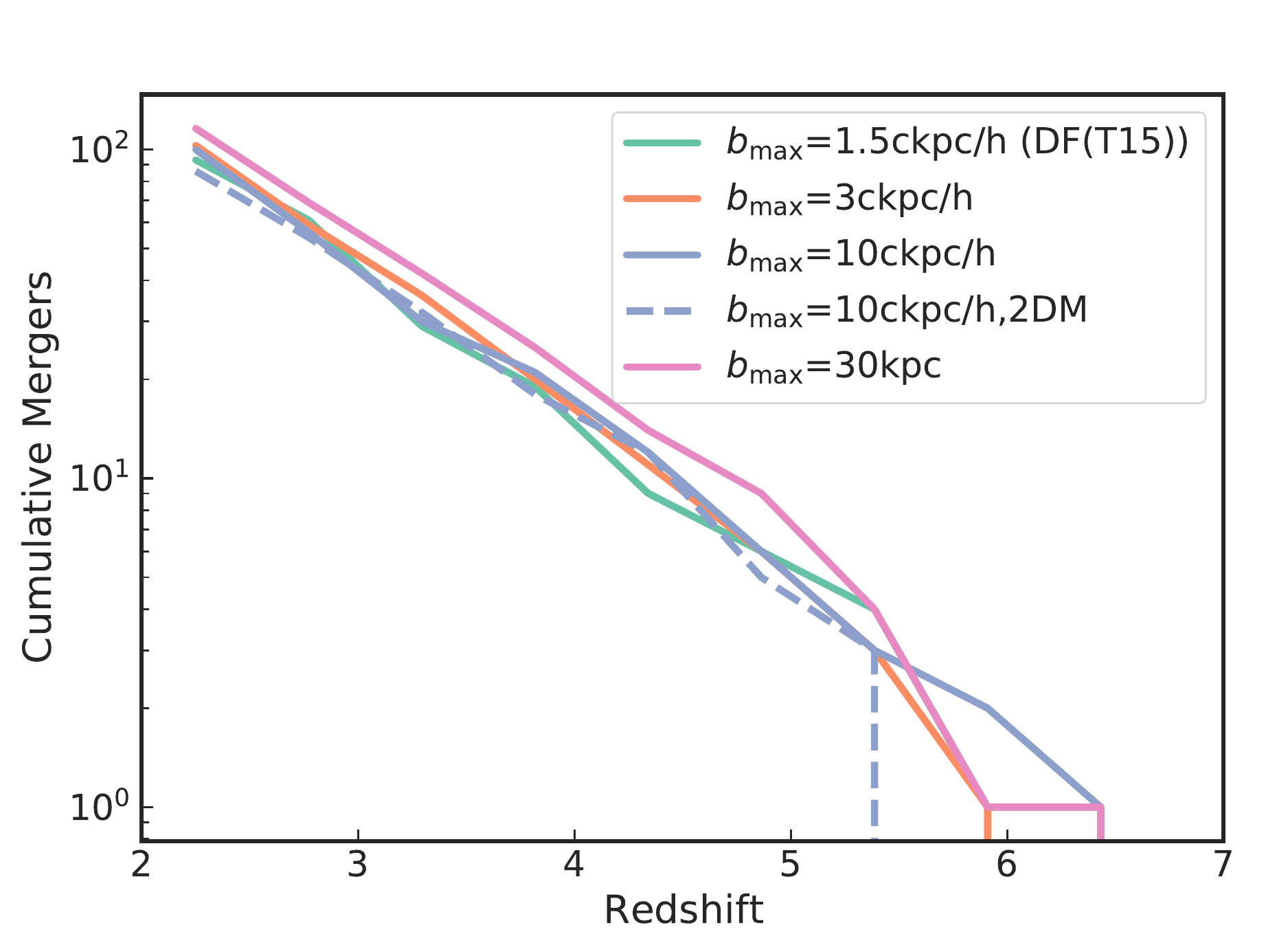}

\caption{The cumulative merger rates for different values of $b_{\rm max}$, in the $L_{\rm  box}$=15 Mpc$/h$ simulations. We tested $b_{\rm max}$ values of 3 ckpc$/h$,10 ckpc$/h$ and 30 kpc, and the difference in the cumulative merger rate is less than 10\%. The difference between the DF(fid) models and the DF(T15) model with $b_{\rm max}$=1.5 ckpc is also very small. Hence, although different choices of $b_{\rm max}$ changes the magnitude of the dynamical friction, it does not affect the merger rate predictions significantly.}
\label{fig:merger_bmax}
\end{figure}

For the merger rate predictions in \ref{sec:L35}, we use the DF+Drag model with $b_{\rm max}$=10 ckpc$/h$ and $M_{\rm dyn,seed} = 4M_{\rm DM}$. In this section, we will show that the merger rate prediction is not sensitive to the choice of these two parameters, and hence our prediction is relatively robust against parameter variations within a reasonable range.

Figure \ref{fig:merger_bmax} shows the cumulative merger rates for different values of $b_{\rm max}$ in the $L_{\rm  box}$=15 Mpc$/h$ simulations. We tested $b_{\rm max}$ values of 3 ckpc$/h$,10 ckpc$/h$ and 30 kpc, and the difference in the cumulative merger rate is less than 10\%. The difference between the DF(T15) models and the DF(fid) model with $b_{\rm max}$=1.5 ckpc is also very small. Hence, although different choices of $b_{\rm max}$ changes the magnitude of the dynamical friction, it does not affect the merger rate predictions significantly.

We also test a lower value of $M_{\rm dyn, seed}=2M_{\rm DM}$ using the $L_{\rm  box}$=15 Mpc$/h$ simulation. The resulting cumulative merger rate prediction is also shown in Figure \ref{fig:merger_bmax}. Compared with the similar run with $M_{\rm dyn, seed}=4M_{\rm DM}$, the earliest merger is slightly postponed, but the cumulative rate at $z\sim 2$ has very little difference. Therefore, even though for the predictions in Section \ref{sec:L35} we have chosen a particular set of parameter values, changing those parameters would not affect the result significantly given the larger effects of other factors such as the resolution and seeding.


\bsp	
\label{lastpage}
\end{document}